\documentclass[final,5p,times,twocolumn,authoryear,nofootinbib]{revtex4-2}

\usepackage{amssymb}
\usepackage{lipsum}
\usepackage{mathtools}
\usepackage{float}
\usepackage{breqn}
\usepackage{xcolor}
\usepackage[normalem]{ulem}

\newcommand{\AEI}{\affiliation{Max Planck Institute for Gravitational Physics (Albert Einstein Institute), Am M\"uhlenberg 1, Potsdam 14476, Germany}}
\newcommand{\Maryland}{\affiliation{Department of Physics, University of Maryland, College Park, MD 20742, USA}}

\begin{document}
\title{Quasinormal modes of slowly-spinning horizonless compact objects}

\author{M. V. S. Saketh}
\email{msaketh@aei.mpg.de}
\Maryland
\AEI

\author{Elisa Maggio}
\email{elisa.maggio@aei.mpg.de}
\AEI

\begin{abstract}

One of the main predictions of general relativity is the existence of black holes featuring a horizon beyond which nothing can escape.
Gravitational waves from the remnants of compact binary coalescences have the potential to probe new physics close to the black hole horizons.
This prospect is of particular interest given several quantum-gravity models that predict the presence of horizonless and singularity-free compact objects.
The membrane paradigm is a generic framework that allows one to parametrise the interior of compact objects in terms of the properties of a fictitious fluid located at the object's radius.
It has been used to derive the quasinormal mode spectrum of static horizonless compact objects. Extending the membrane paradigm to rotating objects is crucial to constrain the properties of the spinning merger remnants.
In this work, we extend the membrane paradigm to linear order in spin and use it to analyse the relationships between the quasi-normal modes, the object's reflectivity, and the membrane parameters.  We find a breaking of isospectrality between axial and polar modes when the object is partially reflecting or the compactness differs from the black hole case.
We also find that in reflective ultracompact objects some of the modes tend towards instability as the spin increases. Finally, we show that the spin enhances the deviations from the black-hole quasinormal mode spectrum as the compactness decreases. 
This implies that spinning
horizonless compact objects may be more easily differentiated than nonspinning ones in the prompt ringdown. 
\end{abstract}

\maketitle

\section{Introduction}
\label{introduction}

Gravitational waves (GWs) provide a unique opportunity to test gravity and investigate the nature of compact objects. 
The ground-based detectors LIGO and Virgo have detected ninety GW candidates from the coalescence of compact binaries up to the end of their third observing run~\cite{LIGOScientific:2018mvr,LIGOScientific:2020ibl,LIGOScientific:2021usb,LIGOScientific:2021djp}.
Some important discoveries include the first observation of a black-hole (BH) coalescence~\cite{LIGOScientific:2016aoc}, the multi-messenger observation of a binary-neutron-star merger~\cite{LIGOScientific:2017vwq}, and an intermediate-mass BH~\cite{LIGOScientific:2020iuh}.
Parametrised tests of general relativity (GR) can constrain the degree to which the data agree with GR by introducing fractional deviations to the inspiral and postmerger stages of the gravitational waveform~\cite{Li:2011cg,Agathos:2013upa,Ghosh:2016qgn,Ghosh:2017gfp,Brito:2018rfr,Carullo:2019flw,Ghosh:2021mrv,Isi:2021iql,Mehta:2022pcn,Maggio:2022hre,Gennari:2023gmx}.
No evidence in support of physics beyond GR has been found by the LIGO-Virgo-KAGRA collaboration, and increasingly stringent limits have been set on GW data~\cite{LIGOScientific:2016lio,LIGOScientific:2019fpa,LIGOScientific:2020tif,LIGOScientific:2021sio}.

GWs offer the promising prospect of testing one of the main predictions of GR, namely the presence of horizons in BHs.
On the theoretical side, BHs feature horizons beyond which nothing~--~not even light~--~can escape.
Moreover, BHs contain a curvature singularity with infinite tidal forces where Einstein equations break down.
In the semiclassical approximation, when a massless field is quantized in the Schwarzschild background, BHs are thermodynamically unstable and radiate a thermal spectrum at the Hawking temperature~\cite{Hawking:1974rv}. They are characterised by the information-loss paradox~\cite{Mathur:2009hf} and their entropy is far over the one of a typical stellar progenitor~\cite{Bekenstein:1973ur,Hawking:1976de}.

Several models of horizonless and singularity-free compact objects have been proposed to address the BH paradoxes~\cite{Giudice:2016zpa,Cardoso:2017cqb,Cardoso:2019rvt,Maggio:2021ans}. New physics can prevent the formation of horizons in quantum-gravity extensions of GR (e.g., fuzzballs~\cite{Mathur:2005zp,Bena:2007kg,Balasubramanian:2008da}, gravastars~\cite{Mazur:2004fk}, nonlocal stars~\cite{Buoninfante:2018xif,Buoninfante:2019swn}) and in GR with the presence of dark matter and exotic fields (e.g., boson stars~\cite{Kaup:1968zz,Ruffini:1969qy,Seidel:1991zh,Liebling:2012fv,Brito:2015pxa}, wormholes~\cite{Morris:1988cz,Visser:1995cc,Damour:2007ap}). Horizonless compact objects can mimic the observational features of BHs~--~ since they can be as compact as BHs~--~and  quantify the existence of horizons in astrophysical sources.
Horizonless compact objects can deviate from BHs for their compactness, i.e., the inverse of the effective radius in units of the total mass,  and reflectivity, which is generically complex and frequency-dependent and differs from the totally absorbing BH case.

The absence of horizons in compact objects affects the gravitational waveform in the inspiral and postinspiral stages.
In the inspiral, the energy flux at the radius of horizonless compact objects differs from the energy flux at the horizon of BHs~\cite{Datta:2019epe,Mukherjee:2022wws,Saketh:2022xjb}.
Furthermore, resonances are excited when the orbital frequency matches the characteristic oscillation frequencies of the inspiralling compact objects~\cite{Cardoso:2019nis,Asali:2020wup,Maggio:2021uge}.
Also, the tidal Love numbers, which measure the deformability of a compact object when immersed in an external tidal field, are  nonzero for horizonless compact objects~\cite{Pani:2015tga,Uchikata:2016qku,Cardoso:2017cfl,Sennett:2017etc,Raposo:2018rjn,Nair:2022xfm,Chakraborty:2023zed}. 
During the ringdown, the quasinormal mode (QNM) frequencies of the perturbed remnant deviate from the BH QNM spectrum in the absence of horizons. If the remnant is an ultracompact horizonless object, the prompt ringdown signal would be nearly the same as the one emitted by a BH since it is excited at the light ring~\cite{Cardoso:2016rao,Cardoso:2016oxy}. Additional signal would be emitted in the form of GW echoes at late times due to the absence of horizons in the remnant~\cite{Abedi:2016hgu,Wang:2019rcf,Mark:2017dnq,Testa:2018bzd,Maggio:2019zyv,Xin:2021zir,Ma:2022xmp,Vellucci:2022hpl,Siemonsen:2024snb}. The QNM spectrum would also deviate from the BH one as a function of the compactness of the object~\cite{Maggio:2020jml}.
Future observatories, such as the Einstein Telescope, Cosmic Explorer and LISA, will allow us to detect or rule out the imprints of horizonless compact objects in the gravitational waveform with unprecedented accuracy~\cite{Branchesi:2023mws,LISA:2022kgy,Cardoso:2017cfl,Maselli:2017cmm,Piovano:2022ojl,Barsanti:2022vvl}.

Here, we study the behaviour of generic horizonless compact objects to external gravitational perturbations to distinguish them from BHs.
It is useful and convenient to incorporate the study of various compact objects into a single framework. This can also lead to an understanding of the relations between the parameters describing compact objects~--~such as compactness and reflectivity~--~and their GW observables~--~such as QNMs and tidal Love numbers.
One way to effectively parametrise a generic compact object is by using the membrane paradigm~\cite{Damour:1982,Thorne:1986iy,Price:1986yy}. The membrane paradigm is a theoretical framework originally developed to describe BHs, according to which the BH interior can be replaced by a three-dimensional membrane located at the event horizon. The membrane is made of a viscous fluid whose density, pressure and viscosity are fixed by the Israel-Darmois junction 
conditions~\cite{Darmois:1927,Israel:1966rt,VisserBook}.

The membrane paradigm was recently applied to compact objects other than BHs by locating the membrane at the object's radius~\cite{Abedi:2020ujo,Maggio:2020jml,Chakraborty:2023zed}. The fluid parameters may be freely varied to parametrise different models of compact objects, and only a specific choice of these parameters describes BHs.

The membrane paradigm was first used to 
derive the QNM spectrum of 
generic nonspinning compact objects in Ref.~\cite{Maggio:2020jml}, along with a comparison with the measurement accuracy of the fundamental QNM of GW150914~\cite{LIGOScientific:2016lio}. However, due to the initial orbital angular momentum of the binary, the final merged remnant is expected to be spinning~\cite{LIGOScientific:2018mvr, LIGOScientific:2020ibl, LIGOScientific:2021usb, LIGOScientific:2021djp}. This necessitates the importance of incorporating the spin into the membrane paradigm for generic compact objects.

In this work, we present an extension of the membrane paradigm for generic compact objects to the linear order in spin. In Sec.~\ref{membrane}, we present the theoretical details involved in constructing the membrane paradigm to linear order in spin. Particularly, we discuss the background metric, the linear perturbations, and outline the derivation of the boundary conditions that the metric perturbations must obey at the membrane surface from the Israel-Damour junction conditions. 
In Sec.~\ref{observe}, we analyse the reflectivity of compact objects as a function of the fluid viscous parameters. We also show the resolution to an apparent paradox that arises in the linear-in-spin approximation associated with the superradiant scattering. Furthermore, we derive the QNM frequencies in the BH limit, and compare them with known Kerr QNM spectrum to quantify the upper limit (in spin) to which the linear-in-spin approximation may be trusted. Finally, we compute the QNM frequencies of generic horizonless compact objects as a function of the compactness and the fluid viscous parameters. We discuss the breaking of isospectrality between axial and polar modes, and  potential constraints that can be put on the object's compactness based on the current measurement accuracies of the fundamental QNM. In Sec.~\ref{end}, we summarise the results and conclude with a brief discussion of the related future extensions.
We assume $G=c=1$ units throughout.

\section{The membrane paradigm}
\label{membrane}

The membrane paradigm allows one to derive the QNM spectrum of compact objects with a given exterior geometry and different interior solutions. In its standard formulation, a static observer can replace the interior of a compact object with a fictitious membrane located at the boundary of the object. 
The Israel-Darmois junction conditions set the properties of the membrane as~\cite{Darmois:1927,Israel:1966rt}
\begin{equation}
 [[K_{ab}-K h_{ab}]] = -8\pi T_{ab}\,, \qquad [[h_{ab}]]=0\,, \label{junction}
\end{equation}
where $h_{ab}$ is the induced metric on the membrane, $K_{ab}$ is the extrinsic curvature, $K=K_{ab}h^{ab}$, and $T_{ab}$ is the stress-energy tensor of the fluid on the membrane. The notation $[[...]]=(...)^+-(...)^-$ denotes the jump of a quantity across the membrane, where $\mathcal{M}^+$ and $\mathcal{M}^-$ are the exterior and interior spacetimes to the membrane, and $a,b$ are the indices of a 3-D coordinate system laid on the membrane.  

The fictitious membrane is such that the extrinsic curvature of the interior spacetime vanishes, i.e., $K_{ab}^-=0$. This constraint allows one to map different geometries for the interior spacetime into the properties of the fictitious membrane.
The Israel-Darmois junction conditions impose that the membrane is a viscous fluid with stress-energy tensor~\cite{Thorne:1986iy}
\begin{alignat}{3}
T_{ab} = \rho u_{a}u_{b} + (p-\zeta \Theta)\gamma_{ab} - 2 \eta\sigma_{ab} \,,
\label{stress}
\end{alignat}
where $\rho$ is the surface energy density, $p$ is the surface pressure, $\zeta$ and $\eta$ are the bulk and shear viscosities, $u_a$ is the 3-velocity of the fluid, $\Theta=u^a_{;a}$ is the expansion, 
$\gamma_{ab}=h_{ab}+u_au_b$ is the projector tensor, $\sigma_{ab} = \frac{1}{2} \left( u_{a;c} \gamma_b^c + u_{b;c} \gamma_a^c - \Theta \gamma_{ab} \right)$ is the shear tensor, and the semicolon is a covariant derivative with the induced metric.

The membrane paradigm was originally developed to describe perturbed BHs~\cite{Damour:1982,Thorne:1986iy,Price:1986yy,Abedi:2020ujo} in terms of the shear and bulk viscosities of a  fictitious viscous fluid located at the horizon, where
\begin{equation}\label{viscosityBH}
\eta_{\rm{BH}}=\frac{1}{16\pi} \,, \quad \zeta_{\rm{BH}}=-\frac{1}{16\pi} \,.
\end{equation}
The membrane paradigm was then extended to perturbed horizonless compact objects with a Schwarzschild exterior~\cite{Maggio:2020jml}, where the fictitious fluid is located at the object's radius, and the shear and bulk viscosity are generically complex and frequency-dependent. For each model of horizonless compact object, the shear and bulk viscosity are uniquely determined and are related to the reflective properties of the object.

\subsection{Background metric}

Motivated by the fact that the remnants of compact binary coalescences have large values of the spin~\cite{LIGOScientific:2018mvr, LIGOScientific:2020ibl, LIGOScientific:2021usb, LIGOScientific:2021djp}, here we extend the membrane paradigm to spinning horizonless compact objects which are generally described by the Kerr metric at the first order in the spin.  Our approach should be considered as a first step towards the inclusion of spin effects into compact bodies, as performed in literature for compact objects beyond Kerr and in modified theories of gravity~\cite{Pani:2012bp,Pani:2013ija,Uchikata:2015yma,Pierini:2021jxd,Wagle:2021tam}. The linear-in-spin approximation is also useful for understanding the new features and conceptual challenges that arise with the inclusion of spin. The extension to higher orders in the spin is left for future work following Refs.~\cite{Hartle:1968si,Pierini:2022eim,Wagle:2023fwl}
It is worth mentioning that the Kerr geometry does not necessarily describe the vacuum region outside a spinning object due to the absence of Birkhoff’s theorem beyond spherical symmetry.
Here, we assume that GR works sufficiently well on the exterior of the compact object, and no specific model is considered for the object's interior.  This assumption is valid in modified theories of gravity  where the corrections to the metric are suppressed by powers of $\ell_P/r_0 \ll 1$, being $r_0$ the object's radius and $\ell_P$ the Planck length or the scale of new physics~\cite{Berti:2015itd}.

The background is the Kerr metric at the linear order in spin, which in Boyer-Lindquist coordinates is given by~\cite{Kojima:1992ie}
\begin{equation}\label{metric}
ds^2
-f(r) dt^2 + \frac{1}{f(r)}dr^2+ r^2 d\Sigma^2
-4 \frac{J}{r}\sin^2 \theta dtd\phi \,,
\end{equation}
where $f(r)=1-2M/r$, $d\Sigma^2 = d\theta^2+\sin^2 \theta d\phi^2$, $J=aM$ is the angular momentum, and $M$ is the total mass of the compact object. For future convenience, we also define  $\chi=a/M$ as the dimensionless spin. Note that for BHs, the magnitude of the spin parameter is $|\chi|\leq 1$ to prevent naked singularities, but no such constraint exists for generic compact objects. A generic spinning body is also characterized by the spin-induced multipole moments, which are relevant from the second order in spin~\cite{Vines:2016qwa,Saketh:2022wap}. For the Kerr metric, all the multipole moments are known as unique functions of $M$ and $J$. In this work,  we truncate to linear order in spin and thus remove contribution of spin-induced multipole moments to the metric. Moreover, the exterior spacetime has no ergoregion, i.e., the region outside the horizon where a static observer cannot exist and is forced to corotate with the compact object.

From the linear-in-spin metric given above, the BH horizon is located at $r_+=2M$. We analyse horizonless compact objects whose radius is located at
\begin{alignat}{3}\label{r0}
r_0 = r_+ (1+\epsilon) \,,
\end{alignat}
where $\epsilon>0$, particularly
$\epsilon = \mathcal{O}(10^{-40})$ for compact objects with Planckian corrections at the horizon
scale, and $\epsilon = \mathcal{O}(1)$ for objects whose compactness is comparable to that of
a neutron star.

In the unperturbed case, the fictitious membrane is located at the object's radius, and the normal vector to the membrane has components 
\begin{equation}
n_t=0, \ n_r=1/\sqrt{f(r)}, \ n_\theta=0, \ n_\phi=0 \,.
\end{equation}
The Israel-Darmois junction conditions in Eq.~\eqref{junction} determine the density, pressure, and 3-velocity of the viscous fluid as   
\begin{alignat}{3}\label{fluid0}
\rho_0&=-\frac{\sqrt{f(r_0)}}{4\pi r_0},~ p_0 =\frac{1+f(r_0)}{16\pi \nonumber r_0\sqrt{f(r_0)}}, \\ 
u^a_0 &= \left(\frac{1}{\sqrt{f(r_0)}},0,\frac{2 J}{r_0^3 \left(1-3f(r_0) \right) \sqrt{f(r_0)}}\right).
\end{alignat}
The detailed derivation of the membrane parameters is given in Appendix~\ref{app:background}.
As expected, the presence of the angular momentum induces a non-zero azimuthal velocity to the fluid. 

\subsection{Linear perturbations}

Let us now perturb the background metric as
\begin{equation}
    g_{\mu\nu}=g_{\mu\nu}^{0} + \epsilon_p \delta g_{\mu\nu} \,,
\end{equation}
where $g_{\mu \nu}^{0}$ is the metric in Eq.~\eqref{metric} and $\epsilon_p$ is a small parameter controlling the strength of perturbation. The metric perturbation $\delta g_{\mu\nu}$ can be split into two parity sectors as~\cite{Regge:1957td,Zerilli:1970se}
\begin{equation}\label{met_pert}
    \delta g_{\mu\nu} = \delta g_{\mu\nu}^{\rm{axial}} + \delta g_{\mu\nu}^{\rm{polar}} \,.
\end{equation}
In the absence of spin, the two sectors evolve independently without mixing. However, in the presence of spin, they mix in the field equations and the boundary conditions. In the slow-rotation limit, it can be shown that the coupling terms between the axial and polar sectors can be dropped for the computation of th QNM frequencies to linear order in spin~\cite{Pani:2013pma}.
After decomposing the perturbation into spherical harmonic basis, the axial and polar perturbations in the Regge-Wheeler gauge to linear order in spin are given by~\cite{Kojima:1992ie,Pani:2013pma}
\begin{eqnarray}
& \delta g_{\mu\nu}^{\text{axial}}  dx^\mu dx^\nu
 = \sum_{\ell m} 2 \Big\{ h_{0}^{\ell m}(r,t) \Big[S_{\theta}^{ \ell m}(\theta,\phi) d\theta + S_{\phi}^{ \ell m}(\theta,\phi) \nonumber \\
 &d\phi\Big]dt + h_{1}^{ \ell m}(r,t) \Big[S_{\theta}^{ \ell m}(\theta,\phi) d\theta + S_{\phi}^{ \ell m}(\theta,\phi)d\phi \Big]dr \Big\} \,,
\end{eqnarray}
\begin{eqnarray}
& \delta g_{\mu\nu}^{\text{polar}} {dx^\mu dx^\nu} = \sum_{\ell m}   \Big[H_0^{ \ell m}(r,t) dt^2 + H_2^{ \ell m}(r,t) dr^2  \nonumber \\&+ r^2 K^{ \ell m}(r,t) d\Sigma^2  +2  H_1^{ \ell m}(r,t)dt dr \Big] Y^{\ell m}(\theta,\phi) \,,
\end{eqnarray}
where $S_{\theta}^{ \ell m}(\theta,\phi) =-\partial_{\phi}Y_{\ell m}(\theta,\phi)/\sin\theta$, $S_{\phi}^{ \ell m}(\theta,\phi) =\sin\theta \partial_{\theta}Y_{\ell m}(\theta,\phi)$, $Y_{\ell m}(\theta,\phi)$ are the tensor spherical harmonics, and $\ell$ and $m$ are the angular and azimuthal numbers of the perturbation.  In the following, we will omit the dependence of the radial and angular functions on the angular and azimuthal numbers 
for brevity of notation.
The time dependence can be factorised out as
\begin{equation}\label{h0time}
    h_0(r,t) = h_0(r) e^{-i \omega t} \,,
\end{equation}
where the same holds for $h_1(r,t)$, $H_0(r,t)$, $H_1(r,t)$, $H_2(r,t)$, $K(r,t)$.

In response to the metric perturbation, the density, pressure and 3-velocity of the membrane stress-energy tensor are perturbed as
\begin{eqnarray}
    \rho &=& \rho_0 + \epsilon_p \delta \rho(\theta,\phi) e^{-i \omega t} \,, \nonumber \\
    p &=& p_0 + \epsilon_p \delta p(\theta,\phi) e^{-i \omega t} \,, \\
    u^a &=& u^a_0 + \epsilon_p \delta u^a(\theta,\phi) e^{-i \omega t} \,, \nonumber
\end{eqnarray}
where $\rho_0$, $p_0$ and $u^a_0$ are given in Eq.~\eqref{fluid0}, and their perturbations can be decomposed in the spherical harmonic basis.
In the spinless case ($J=0$) the axial perturbations couple with the azimuthal component of the velocity perturbation of the fluid, whereas the polar perturbations couple with the density perturbation, the pressure perturbation, and the radial and polar deformations of the membrane location defined in Eq.~(\ref{appeq : rad_sft}). 
At the first order in spin, the perturbations with a given parity and angular number $\ell$ are coupled to those with opposite parity and angular number $\ell \pm 1$.
However, it can be shown that the couplings to the $\ell \pm 1$ terms do not contribute to the QNM spectrum to the linear order in spin~\cite{Pani:2013pma}. Thus, we neglect these terms in the following, and work with axial and polar modes separately. See Appendix~\ref{app : mem bc} for the derivation of the equations relating the metric perturbations to the fluid perturbations of the membrane.

Outside the membrane, the Einstein equations governing the metric perturbations can be reduced to two Schrödinger-like equations, namely the Regge-Wheeler and Zerilli equations describing axial and polar perturbations, respectively~\cite{Regge:1957td,Zerilli:1970se}. They are of the form: 
\begin{equation}\label{schrodingereq}
    \frac{d^2 \psi(r)}{d r_*^2} + \left[ \omega^2 - \frac{4 m J \omega }{r^3} - V(r)\right] \psi(r) =0 \,,
\end{equation}
where the Regge-Wheeler effective potential at linear order in the spin is~\cite{Kojima:1992ie,Pani:2013pma}
\begin{equation}
    V_{\rm{RW}} = f(r) \left[ \frac{\ell(\ell+1)}{r^2}- \frac{6 M}{r^3} + \frac{24 J m (3 r-7M) }{\ell(\ell+1) \omega r^6 } \right] \,, 
\end{equation}
and the Zerilli effective potential is given in Eq.~\eqref{appeq : Zer pot}, where the tortoise coordinate is defined as $dr_*/dr = 1/f(r)$. The Regge-Wheeler function $\psi_{\text{RW}}(r)$ is related to the metric perturbation function $h_1(r)$ as
\begin{alignat}{3}
\psi_{\text{RW}}(r) = \left(1-\frac{2 m J}{r^3 \omega}\right)^{-1}\frac{f(r)}{r} h_1(r) \,,
\end{alignat}
and $h_0(r)$ is related to $h_1(r)$ via Eq.~(\ref{app eq h1toh0}). Similarly, the Zerilli function $\psi_{\text{Z}}(r)$ can be written as a linear combination of $H_0(r)$, $H_1(r)$, $H_2(r)$ and $K(r)$ as in Eqs.~\eqref{appeq : Zer func1}-\eqref{H0H1K} at linear order in spin. 

\subsection{Boundary conditions}

By imposing boundary conditions at infinity and the radius of the compact object, Eq.~\eqref{schrodingereq} defines an eigenvalue problem.
From the junction condition in Eq.~\eqref{junction}, one finds a constraint on the metric perturbation at the membrane surface, i.e., a boundary condition at the radius of the compact object.
Different models of horizonless compact objects are described by different boundary conditions, which depend on the membrane parameters such as the shear and bulk viscosities of the fluid. 
The boundary condition at membrane surface  for axial perturbations turns out to be
\begin{alignat}{3}
\label{eq : axial main bc}
& \frac{1}{\psi_{\text{RW}}(r)}\frac{d\psi_{\text{RW}}(r)}{dr_*}\Big|_{r=r_0}
= - \frac{i \omega}{\nu} + \frac{M y^2 V_{\text{RW}}(r_0)}{2 w (3w-y)} + \nonumber \\&\nonumber \frac{m \chi}{2 \ell (\ell+1) M y^5 (y-3 w)^2\nu}\Big\{w^3 \Big[36 w^2 y \Bigl(i \left(\ell^2+\ell+9\right) y+ \\&  16 \nu+\nu  y^2\Bigl)+6 w y^2 \Bigl(\left(\ell^2+\ell-7\right) \nu  y^2 -2 i y \left(2 \ell^2+2 \ell+13\right) \nonumber \\& -42 \nu \Bigl)\nonumber +y^3 \Bigl(-3 \left(\ell^2+\ell-4\right) \nu  y^2+4 i \left(\ell^2+\ell+6\right) y+ \nonumber \\ & 36 \nu \Bigl)-216 i w^3 \left(y-2 i \nu\right)\Big]\nonumber  -3 w^3 (2w-y) \left(6w-\ell(\ell+1)y\right) \\
&\Bigl[\left(4  w- 2 y + y^3\right)\nu -2 i y(3w-y) \Bigl]\Big\},
\end{alignat}
where $\nu=16 \pi \eta$, $y=\omega r_0$, and $w=M\omega$.
The polar boundary condition turns out to be
\begin{alignat}{3}
\label{eq : polar main bc}
\frac{1}{\psi_{\text{Z}}(r)}\frac{d\psi_{\text{Z}}(r)}{dr_*}\Big|_{r=r_0} &= -16 \pi \eta i \omega +  \nonumber G(r_0,\omega,\eta,\zeta) \\&+ m \chi H(r_0,\omega,\eta,\zeta) \,,
\end{alignat}
where $G(r_0,\omega,\eta,\zeta)$ is given  in Ref.~\cite{Maggio:2020jml}, and the supplemental Mathematica notebook in~\cite{githublink} contains the complete expression for  the polar boundary condition including $H(r_0,\omega,\eta,\zeta)$. The detailed derivation of the above boundary conditions starting from the junction conditions in Eq.~(\ref{junction}) is provided in Appendix~\ref{app : mem bc}.

Note that the above boundary conditions are invariant under the transformation $(m\rightarrow -m$,~$\chi\rightarrow -\chi)$. It is also worth noting that the axial boundary condition does not depend on the bulk viscosity $\zeta$ of the fluid membrane. 

The BH boundary condition is recovered in the $\epsilon\rightarrow 0$ limit, for which $V(r_0)=0$ and $y=2w$, and by setting the membrane shear and bulk viscosities to BH values as in Eq.~\eqref{viscosityBH}.
In this limit, the axial and polar boundary conditions in Eqs.~(\ref{eq : axial main bc}) and~(\ref{eq : polar main bc}) reduce to
\begin{alignat}{3}\label{horizonasympt}
\frac{d\psi_{\rm{RW}}(r)/dr_*}{\psi_{\rm{RW}}(r)}\Big|_{r=r_+}=\frac{d\psi_{\rm{Z}}(r)/dr_*}{\psi_{\rm{Z}}(r)}\Big|_{r=r_+}= - i \tilde{\omega} \,,
\end{alignat}
where $\tilde{\omega}=\omega-m\Omega_H$ and $\Omega_H=J/(4 M^3)$ is the horizon angular velocity. This is consistent with the expected BH boundary condition of a purely ingoing perturbation near the horizon, i.e., 
\begin{equation}
    \psi_{\rm{RW}}(r)\sim \psi_{\rm{Z}}(r) \sim e^{-i \tilde{\omega} r_*} \,, \quad \text{as} \ r_* \to -\infty \,.
\end{equation}
Finally, we check that in the spinless limit, $\chi\rightarrow 0$, the boundary conditions in Eqs.~(\ref{eq : axial main bc}) and~(\ref{eq : polar main bc}) reduce to those obtained in Ref.~\cite{Maggio:2020jml} for a horizonless compact object with a Schwarzschild exterior.

\section{Reflectivity and quasinormal mode spectrum}
\label{observe}
\subsection{Reflectivity of compact objects}
\label{subsec : obs refl}
\begin{figure}[t]
    \centering
    \includegraphics[width=0.49 \textwidth]{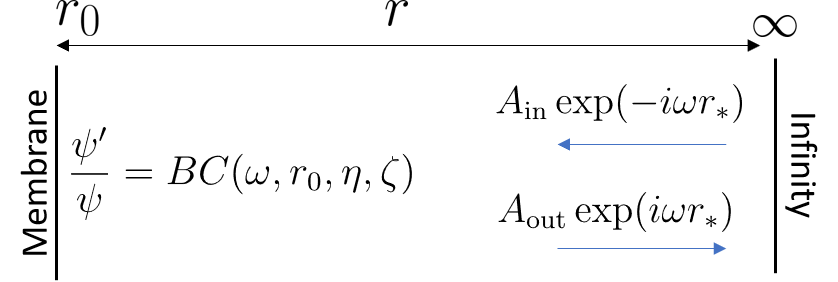}
    \caption{Time-independent scattering of gravitational perturbations off a compact object, which is described by a membrane located at the object's radius. At infinity, the Regge-Wheeler and Zerilli wave functions asymptote towards free-wave solutions. At the membrane surface, the boundary conditions in Eqs.~(\ref{eq : axial main bc}) and (\ref{eq : polar main bc}) are obeyed. The reflectivity of the compact object at infinity is defined as $\mathcal{R}=A_{\mathrm{out}}/A_{\mathrm{in}}$.}
    \label{fig : scatter}
\end{figure}
One way to differentiate between compact objects is based on their reflectivity as measured from infinity. Given an ingoing wave packet sent from infinity in the distant past, 
the reflectivity is defined as the resultant scattered waveform at infinity in the distant future. It is convenient to study a time-independent scattering, where both ingoing and outgoing spherical waves are simultaneously present. This is schematically represented in Fig.~\ref{fig : scatter}, where the ratio of the amplitudes of the outgoing and ingoing waves at infinity defines the reflectivity as
\begin{equation}\label{reflectivity}
    \psi(r) \sim e^{-i \omega r_*} + \mathcal{R} e^{i \omega r_*} \,, \quad \text{as} \ r_* \to \infty \,,
\end{equation}
after the wave interacts with the compact object and is subjected to the boundary conditions in Eqs.~\eqref{eq : axial main bc} and~\eqref{eq : polar main bc} for axial and polar perturbations, respectively.

For non-spinning BHs, 
with a horizon but no ergoregion, 
the absolute value of reflectivity, $|\mathcal{R}|$, is bounded from above by 1. For small frequencies, or equivalently large wavelengths, the wave packets larger than the BH size are only negligibly absorbed,
yielding $|\mathcal{R}|\approx 1$. As the wavelength becomes comparable with the BH size,
absorption is more relevant and $|\mathcal{R}|<1$. 

For Kerr BHs possessing an ergoregion, the reflectivity can  exceed $1$ when the frequency of the perturbation is smaller than the BH horizon angular velocity, i.e., $\omega<m\Omega_H$. This is due to the phenomenon of superradiance, in which the energy associated with the object's rotation may be transmitted to the waves/particles scattered by it. In this case, the BH loses its rotational energy, leading to an amplification of the outgoing radiation. The analytical expression for the reflectivity of Kerr BHs at low frequencies to the next-to-leading order in $M\omega$ was derived in Refs.~\cite{Starobinsky:1973aij,Starobinskij2,PhysRevD.104.124061,Saketh:2022xjb,Dolan:2008kf,Saketh:2023bul}.

For generic compact objects, the reflectivity $|\mathcal{R}|$ is related to the membrane parameters (i.e., shear and bulk viscosity of the fluid) through the boundary conditions in Eqs.~(\ref{eq : axial main bc}) and~(\ref{eq : polar main bc}) for axial and polar perturbations, respectively. The computation of the reflectivity for arbitrary frequencies is done via numerical integration of the perturbation equations in Eq.~\eqref{schrodingereq}  and discussed in Sec.~\ref{sec:nuref}. Some features of the reflectivity may be extracted analytically and are discussed in the next few paragraphs. We will also discuss the limitations of the linear-in-spin approximation in the form of spurious effects, specifically superradiance, leading to $|\mathcal{R}|>1$ when the spin parameter $\chi$ becomes large enough. This is physically not allowed as superradiance requires the presence of an ergoregion, which is absent at the linear order in spin. Nevertheless, this spuriousness is understood in terms of the limits of validity of the linear-in-spin approximation.

For an ultracompact object ($\epsilon \ll 1$), in the limits $r\rightarrow 2M$ and $r\rightarrow \infty$, the Regge-Wheeler and Zerilli equations at the linear order in spin 
reduce to free-wave equations
\begin{alignat}{3}
\frac{d^2\psi(r)}{dr_*^2}+\left[\omega^2-\frac{m \chi \omega}{2M}\right]\psi(r)=0, &\quad r\rightarrow 2M \,, \label{asy1} \\
\frac{d^2\psi(r)}{dr_*^2}+\omega^2\psi(r)=0, &\quad r\rightarrow \infty, \label{asy2}
\end{alignat}
In both these limits, the asymptotic solutions can be written as 
\begin{alignat}{3}
\psi(r) \sim C_{\text{trans}} \ e^{-i\omega'r_*}+C_{\text{ref}} \ e^{i\omega' r_*}, \quad &r \to 2M \,, \label{eq : Wronkshor}\\ 
\psi(r) \sim A_{\text{in}} \ e^{-i\omega r_*} + A_{\text{out}} \ e^{i\omega r_*}, \quad &r \to \infty \,,
\label{eq : Wronks}
\end{alignat}
where $\omega' =\omega\sqrt{1 -m\chi/(2M\omega)}$ and $\omega\in \text{Reals}$. For BHs, the purely ingoing boundary condition at the horizon fixes $C_{\text{ref}}=0$.
In order to provide an analytical description of the reflectivity, we use the following conservation law valid for the solution of a Schrödinger-like equation $\psi''+(\omega^2-V)~\psi=0$ with real potential and frequency:
\begin{alignat}{3}
& \frac{d}{dr_*}\left(\psi' \bar{\psi} - \bar{\psi}'\psi \right) = \psi''\bar{\psi}-\bar{\psi}''\psi  \nonumber = - \tilde{V} \psi \bar{\psi}+\tilde{V} \bar{\psi} \psi=0\\&\implies \left(\psi' \bar{\psi} - \bar{\psi}'\psi\right)\Big|_{r=2M} =\left(\psi' \bar{\psi} - \bar{\psi}'\psi\right)\Big|_{r\to\infty} , 
\label{eq : gencons}
\end{alignat}
where the prime denotes a derivative with respect to the tortoise coordinate, $\bar{\psi}$ is the complex conjugate of $\psi$, and
$\tilde{V}=-\omega^2+4mJ\omega/r^3 + V(r)$. 
Eq.~\eqref{eq : gencons} together with Eqs.~\eqref{asy1},\eqref{asy2} implies in turn the relation
\begin{alignat}{3}
\frac{\text{Re}(\omega')}{\omega}|C_{\text{trans}}|^2 e^{2 \text{Im}(\omega') r_*}= |A_{\text{in}}|^2-|A_{\text{out}}|^2 \,,
\label{eq : conservation}
\end{alignat}
where we note that $\omega'\in \text{Reals}$, provided  $\omega>m \chi/ 2 M$, and furthermore $\omega'>0$. Thus, for sufficiently large frequencies, i.e., $\omega>m \chi/ 2 M$, we have $|A_{\rm{in}}|^2 > |A_{\rm{out}}|^2$ or $|\mathcal{R}|^2<1$ as expected from BHs.  When the frequency is small, i.e., $\omega<m \chi/ 2 M$, $\omega'$ becomes purely imaginary and the right hand side of Eq.~\eqref{eq : conservation} vanishes, yielding $|\mathcal{R}|^2=1$. This is different from the behaviour of Kerr BHs where one observes a superradiant amplification, i.e., $|\mathcal{R}|^2>1$, when $\omega<m\Omega_H$~\cite{Penrose:1971uk,1971JETPL..14..180Z,Blandford:1977ds,Brito:2015oca}. This can be explained by the fact that we are using a linear-in-spin approximation of the Kerr metric, which does not possess an ergoregion and cannot reproduce a superradiant scattering. 

Something interesting happens, however, when we expand $\omega'$ to linear order in spin, i.e., $\omega'  = \omega-m\chi/(4M) + \mathcal{O}(\chi^2)$, which modifies Eq.~(\ref{eq : conservation}) to
\begin{alignat}{3}
\frac{\tilde{\omega}}{\omega}|C_{\text{trans}}|^2 = |A_{\text{in}}|^2-|A_{\text{out}}|^2,
\label{eq : fake superrad}
\end{alignat}
where $\tilde{\omega}$ is defined below Eq.~\eqref{horizonasympt} and $\tilde{\omega} \in \text{Reals}$.
Now, we find a superradiant behaviour when $\omega<m\Omega_H$, which is analogous to the Kerr BH case. 
Nevertheless, since 
we do not expect superradiance at linear order in spin due to the absence of an ergoregion, we  interpret this as a spurious result. Indeed, a consistent study of the slow rotation approximation requires going to at least second order in spin~\cite{Pani:2012bp,Pani:2013pma}. This argument requires extra care when dealing with membranes as their boundary condition is valid up to leading order in spin 
(unlike the expression in Eqs.~\eqref{eq : Wronkshor}). 
We thus interpret any observation of effects related to the presence of an ergoregion as a spurious result. This will become relevant for the analysis of the stability of QNMs as shown in Sec.~\ref{sec:QNMs}.

For the compact objects analysed here, the asymptotic behaviour of the perturbations at infinity in Eq.~(\ref{eq : Wronks}) is still valid, however the asymptotic solution near the object's radius does not have a simple analytical form. The following conservation law holds
\begin{alignat}{3}
\left(\psi' \bar{\psi} - \bar{\psi}'\psi\right)\Big|_{r=r_0} =(\psi' \bar{\psi} - \bar{\psi}'\psi)\Big|_{r=\infty} , 
\label{eq : gencons mem}
\end{alignat}
where $r_0$ is the location of the membrane. We now use the boundary conditions in Eqs.~(\ref{eq : axial main bc}) and~(\ref{eq : polar main bc}) for axial and polar perturbations to rewrite the conservation law as
\begin{alignat}{3}
-|\psi|^2\frac{\text{Im}[\text{BC}(\omega,r_0,\eta,\zeta)]}{i\omega} =|A_{\text{in}}|^2-|A_{\text{out}}|^2\,
\label{eq : gencons mem}
\end{alignat}
where $\text{BC}(\omega,r_0,\eta,\zeta)=\psi'/\psi|_{r=r_0}$ is the boundary condition to be satisfied by the relevant perturbation functions (Regge-Wheeler or Zerilli). In the BH limit, i.e., $\epsilon\rightarrow 0$, $\eta\rightarrow \eta_{\rm{BH}}$ and $\zeta\rightarrow \zeta_{\rm{BH}}$, Eq.~\eqref{eq : gencons mem} reduces to Eq.~(\ref{eq : fake superrad}) as expected.
Since the exterior spacetime is the Kerr metric at the linear order in spin, which does not have an ergoregion, we cannot trust the regime in which $|\mathcal{R}|^2>1$. It is also worth noting that a vanishing of the imaginary part of the boundary condition on the membrane leads to a perfect reflection where $|A_{\rm{out}}|^2 = |A_{\rm{in}}|^2$. This can also be achieved by  setting $\psi=0$ on the membrane, corresponding to a Dirichlet boundary condition.

\subsubsection{Numerical computation of the reflectivity : Methods}

The reflectivity can be computed numerically by integrating the perturbation equations in Eq.~\eqref{schrodingereq} with the boundary conditions represented in Fig.~\ref{fig : scatter}.  
The second-order Regge-Wheeler and Zerilli equations are integrated outwards starting from the membrane radius $r_0$, where the wave function and its first derivative need to be specified at the membrane surface. Among the two unknown functions, one is simply a scale, therefore we can set $\psi(r_0)=1$. Then, $\psi'(r_0)$ is specified by the boundary conditions in Eqs.~(\ref{eq : axial main bc}) and (\ref{eq : polar main bc}) for axial and polar perturbations, respectively.

The equations are numerically integrated  outwards up to a point sufficiently far away (say $r_{\infty}=50M$), which may be regarded as ``infinity''. We check that the result of the integration is numerically stable by changing the numerical value for the infinity.
The asymptotic solutions at infinity are given in Eq.~(\ref{eq : Wronks}). Then, given the (numerically obtained) wavefunction and its first derivative, we compute the reflectivity as
\begin{alignat}{3}
|\mathcal{R}| = \Bigg|\frac{A_{\text{out}}}{A_{\text{in}}}\Bigg| = \Bigg|\frac{i\omega \psi(r_{\infty})+\psi'(r_{\infty})}{i\omega\psi(r_{\infty})-\psi'(r_{\infty})}\Bigg|.
\label{eq : rinftoR}
\end{alignat}
The accuracy of the integration can be improved by making use of series solutions at infinity or close to the horizon in the BH case. This modifies the relation in Eq.~(\ref{eq : rinftoR}) between the wavefunction and the reflectivity at infinity accordingly. However, concerning the computation of the reflectivity, series solutions at infinity do not affect significantly the result. For instance, for a compact object with $\epsilon=10^{-10}$ and $\eta=\eta_{\rm{BH}}$, the absolute value of the reflectivity obtained without asymptotic series solutions at infinity and using a nineteenth-order asymptotic series solution in powers of $1/r$ differ by less than $0.1\%$ at the frequency $M\omega=0.37$. For this reason, we use Eq.~(\ref{eq : rinftoR}) for the computation of the reflectivity, where $\psi(r_\infty)$ and $\psi'(r_\infty)$ are obtained from a direct integration.

\subsubsection{Numerical computation of the reflectivity : Results}
\label{sec:nuref}

\begin{figure}[t]
\centering
\begin{minipage}{0.45\textwidth}
\centering
\includegraphics[width=0.99\textwidth]{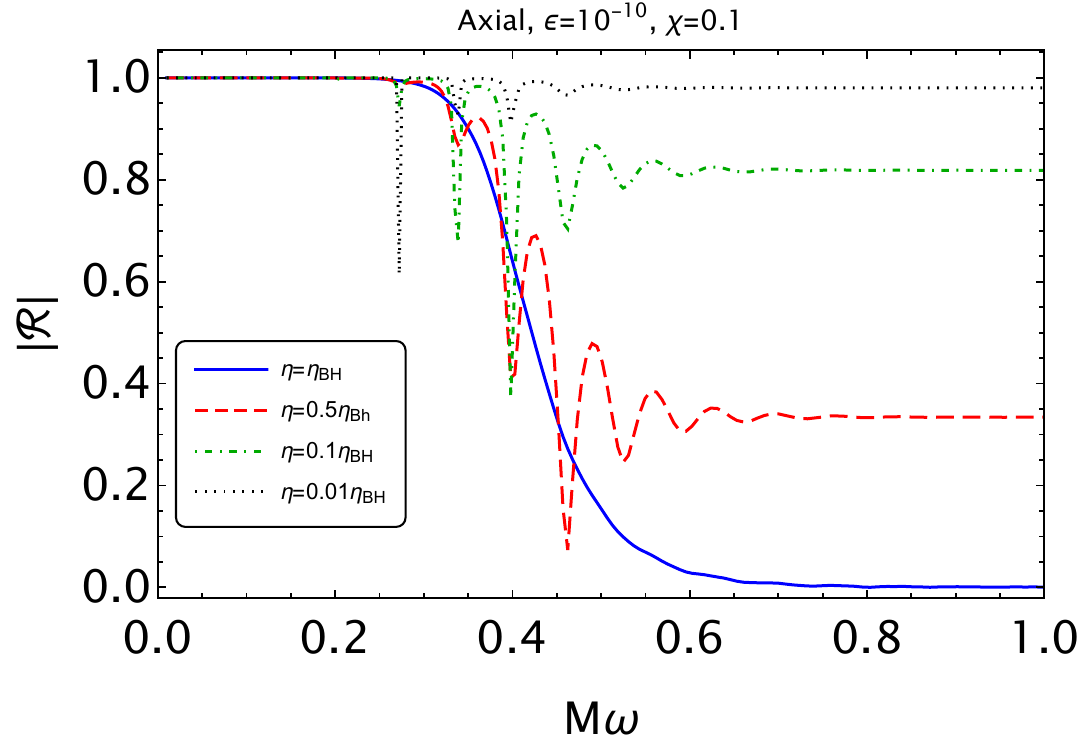}
\end{minipage}\hfill
\begin{minipage}{0.45\textwidth}
\centering
\includegraphics[width=0.99\textwidth]{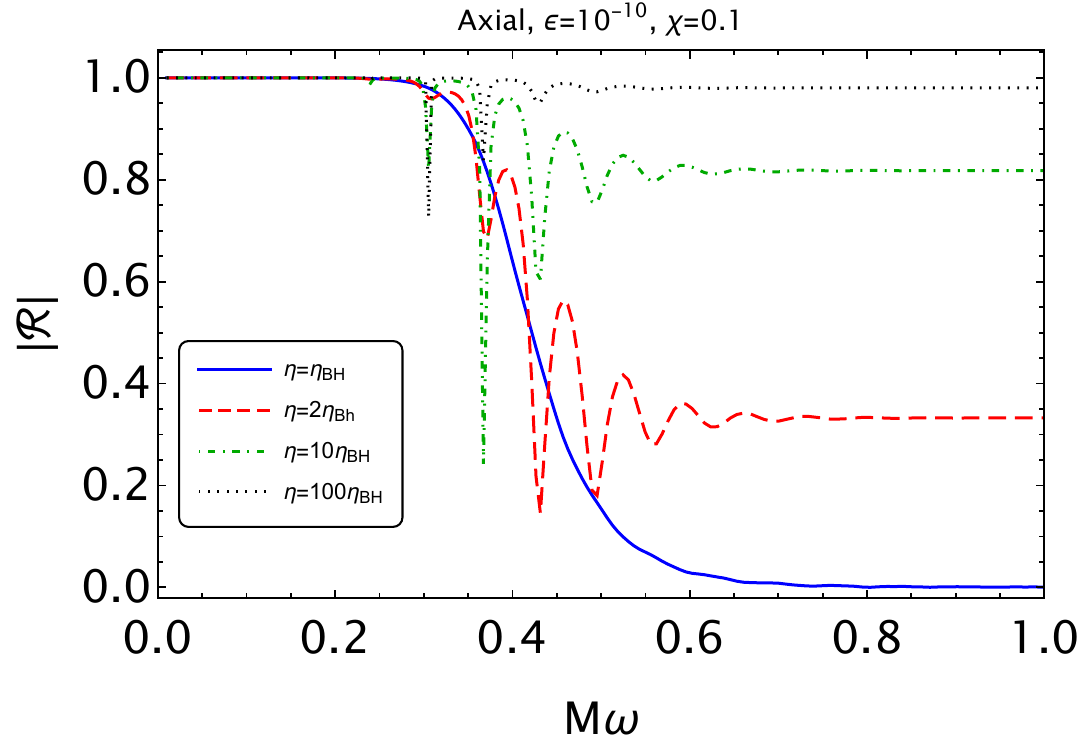}
\end{minipage}
\caption{Reflectivity of a compact object for axial perturbations in terms of the shear viscosity, $\eta$, of a fictitious fluid located at the object's radius, $r_0 = 2M(1+\epsilon)$, with $\epsilon=10^{-10}$, $\chi=0.1$ and $\ell=m=2$. 
The shear viscosity varies as $\eta \in (0,\eta_{\rm{BH}})$ in the top panel and $\eta \in (\eta_{\rm{BH}},\infty)$ in the bottom panel. Note that the limits $\eta \to 0$ and $\eta \to \infty$ lead to a perfectly reflecting compact object, described by Dirichlet and Neumann boundary conditions, respectively. When $\eta \neq \eta_{\rm{BH}}$, the reflectivity has an oscillatory structure, which is related to the QNMs of the object.
}
\label{fig:reflaxial}
\end{figure}

Let us discuss the variation of reflectivity of the compact object with respect to the fluid parameters.
The shear viscosity $\eta$, appearing in both the axial and polar boundary conditions, controls the reflectivity of the object in the ultracompact limit. This can be seen from the $\epsilon\rightarrow 0$ limit of Eqs.~(\ref{eq : axial main bc}) and (\ref{eq : polar main bc}), which is

\begin{alignat}{3}
\label{eq : axial eps0 lim}
\frac{1}{\psi_{\rm{RW}}(r)}\frac{d\psi_{\rm{RW}}(r)}{dr_*}\Big|_{r=r_0} &= - \frac{i}{16 \pi \eta}\Big(\omega-m\frac{\chi}{4M}\Big), \\
\label{eq : polar eps0 lim}
\frac{1}{\psi_{\rm{Z}}(r)}\frac{d\psi_{\rm{Z}}(r)}{dr_*}\Big|_{r=r_0} &= - 16 \pi i  \eta \Big(\omega-m\frac{\chi}{4M}\Big) + \\&  \frac{\left(\ell^2+\ell-1\right)m\left[\left(16\pi\eta\right)^2-1\right]\chi\omega}{2l\left(\ell+1\right)\left(\ell^2+\ell+1\right)} .
\nonumber
\end{alignat}
In the BH limit, $\eta\rightarrow \eta_{\rm{BH}}$, the above equations reduces to the purely ingoing boundary condition as expected from BHs.  In the spinless limit ($\chi=0$), the limit $\eta \to 0$ corresponds to a Dirichlet ($\psi_{\rm RW}=0$) and a Neumann ($\psi'_{\rm{Z}}=0$) boundary condition on axial and polar perturbations, respectively. Whereas the limit $\eta \to \infty$ corresponds to a Dirichlet ($\psi_{\rm Z}=0$) and a Neumann ($\psi'_{\rm{RW}}=0$) boundary condition on polar and axial perturbations, respectively. In both of these  limits, the compact object is purely reflecting at any frequency, i.e., $|\mathcal{R}|=1$. Note that when the spin is turned on ($\chi\neq 0$), the limit $\eta\rightarrow 0$ no longer corresponds to a Neumann boundary condition for polar perturbations. However, this limit still describes a purely reflective compact object with $|\mathcal{R}|=1$. This can be seen from Eq.~(\ref{eq : gencons mem}), where the imaginary part of the boundary condition (i.e., the right hand side of Eq.~(\ref{eq : polar eps0 lim})) vanishes in the limit $\eta\rightarrow 0$ while $\psi_{\rm{Z}}(r_0)$ remains finite.

Fig.~\ref{fig:reflaxial} shows the absolute value of the reflectivity of a compact object with $\epsilon\ll 1$ and different values of the shear viscosity, particularly $\eta \in (0,\eta_{\rm{BH}})$ in the top panel and $\eta \in (\eta_{\rm{BH}},+\infty)$ in the bottom panel. 
We have set $\chi=0.1$ for simplicity. The plots are for axial perturbations, but they look qualitatively similar in the polar case, yielding to the same results. 
The reflectivity profile coincides with the BH one when $\eta=\eta_{\rm{BH}}$, and describes a perfectly reflecting object in the $\eta \to 0$ and $\eta \to +\infty$ limits. The approximate symmetry in the two panels of Fig.~\ref{fig:reflaxial} can be understood analytically in the large-frequency limit, where the absolute value of the reflectivity is invariant under the transformation $\eta/\eta_{\rm{BH}} \rightarrow \eta_{\rm BH}/\eta$, as shown in Eq.~(16) of Ref.~\cite{Maggio:2020jml}. 
An example of this apparent symmetry is provided by the wormhole, which is a purely reflecting object that can be described by  either Dirichlet or Neumann boundary conditions, corresponding to $\eta \to 0$ and $\eta \to \infty$~\cite{Cardoso:2016rao,Maggio:2017ivp,Maggio:2018ivz}.
While the absolute value of the reflectivity may be  similar in the large-frequency limit, the quasinormal mode spectrum is generally different as seen in the differing oscillatory pattern at smaller values of frequency.
Note that the oscillatory structure in reflectivity becomes sharply peaked when $\eta \to 0$ and $\eta \to +\infty$. This behaviour is due to the QNMs of the compact object, where the frequencies at which the dips (or resonances) occur coincide with the real part of the QNMs of the object. Using the analytical formulae for the QNMs of ultracompact objects in the small-frequency regime~\cite{Maggio:2018ivz,Maggio:2021uge}, we check that the distance between subsequent dips (or resonances) is given by
\begin{equation}
   \Delta \omega=\frac{\pi}{|r_*^0|} \sim \frac{\pi}{2 M |\log \epsilon|} \,.
\end{equation}
The resonance width is proportional to the imaginary part of the QNMs~\cite{Maggio:2018ivz,Maggio:2021uge}, which becomes smaller in the perfectly reflecting cases compared to the partially absorbing ones. This explains the sharpness of the resonances in the $\eta\to0$ and $\eta \to \infty$ limits.

\begin{figure}[t]
\centering
\includegraphics[width=0.45\textwidth]{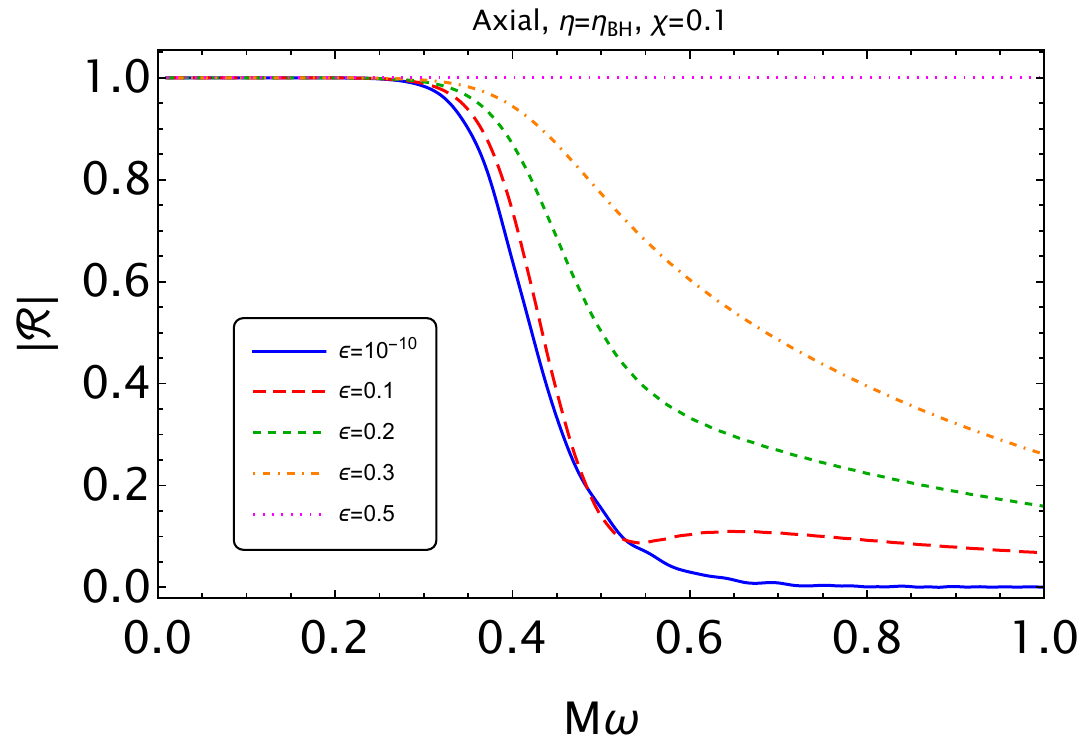}
\includegraphics[width=0.45\textwidth]{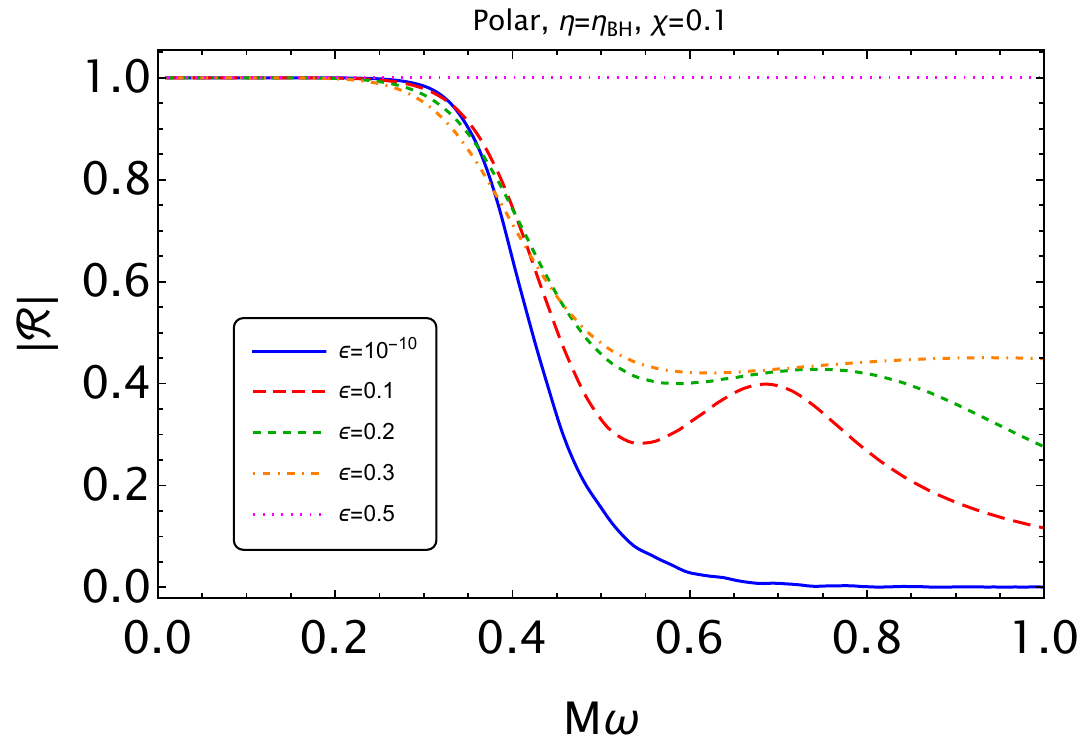}
\caption{
Reflectivity of a compact object for axial (top panel) and polar (bottom panel) perturbations in terms of the shear ($\eta=\eta_{\rm{BH}}$) viscosity of a fictitious fluid located at the object's radius, $r_0 = 2M(1+\epsilon)$, for $\chi=0.1$ and $\ell=m=2$.
The object's compactness varies as $\epsilon \in (10^{-10},0.5)$. Note that the reflectivity approaches 1 as $\epsilon\rightarrow 0.5$, which corresponds to the object's radius approaching the light ring.}
\label{plot : refl epsvary spin0d1}
\end{figure}

Let us now analyse the variation of the reflectivity with the object's compactness for a fixed viscosity. For this purpose, we fix the viscosity in the axial boundary condition in Eq.~(\ref{eq : axial main bc}) to the BH value. The purely ingoing boundary condition is recovered for $\epsilon\rightarrow 0$, as seen from Eq.~(\ref{eq : axial eps0 lim}). However, in the limit $\epsilon\rightarrow 1/2$, corresponding to the membrane radius located at the light ring, $r_0 \rightarrow 3M$, the reflectivity becomes unity as the axial boundary condition tends to the Dirichlet boundary condition, i.e., $\psi_{\rm{RW}}(r_0)=0$. 
This happens regardless of the spin as the second and  third terms in the right hand side of Eq.~(\ref{eq : axial eps0 lim}) are divergent. The top panel of Fig.~\ref{plot : refl epsvary spin0d1} shows the variation of the reflectivity with the object's compactness for axial modes in the slowly spinning case where $\chi=0.1$.
The reflectivity coincides with the BH one when $\eta=\eta_{\rm{BH}}$ and $\epsilon$ is
small enough, and describes a perfectly reflecting object in the $\epsilon \to 1/2$ limit.

In the polar case, interestingly, the boundary condition in the spinless case ($\chi=0$) does not reach the perfectly reflecting limit as $\epsilon\rightarrow 1/2$. Moreover, the reflectivity depends on the choice of the bulk viscosity of the membrane fluid. However, the linear-in-spin term $H(r_0, \omega, \eta, \zeta)$ in Eq.~(\ref{eq : polar main bc}) is divergent in the limit $\epsilon\rightarrow 1/2$, thus changing the behaviour even for a small value of spin. The bottom panel of Fig.~\ref{plot : refl epsvary spin0d1} shows the variation of the reflectivity with the object's compactness for polar modes in the slowly spinning case where $\chi=0.1$.
Note that the small-spin approximation breaks down for axial and polar modes in the limit $\epsilon\rightarrow 1/2$. This is because the linear-in-spin term of the boundary condition scales as $ (\epsilon-1/2)^{-2}$, whereas the spinless term of the boundary condition scales as $(\epsilon-1/2)^{-1}$, as shown in Eq.~(\ref{eq : axial main bc}). 
Nevertheless, due to the boundary condition diverging in the limit $\epsilon\rightarrow 1/2$  both in the nonspinning and spinning cases, for $\epsilon=1/2$ the compact object is purely reflecting. It is important to mention that for a given model of compact object, the viscosities ($\eta$, $\zeta$) are uniquely determined and will depend on the compactness parameter $\epsilon$. 
For simplicity, here we keep the viscosities fixed while varying the compactness (and vice versa). Model-specific studies are left for future work.

\subsection{Quasinormal mode spectrum}
\label{sec:QNMs}

Another important observable, which is useful for differentiating between compact objects, is the QNM spectrum. 
By imposing boundary conditions at infinity and the radius of the compact object, the complex eigenvalues of Eq.~\eqref{schrodingereq} are the QNM frequencies of the system, $\omega = \omega_R + i \omega_I$. 
In our convention, a stable mode has $\omega_I<0$ and corresponds to an exponentially damped sinusoidal signal 
with frequency $f = \omega_R/(2 \pi)$ and damping time $\tau_{\text{damp}}=-1/\omega_I$. Conversely, an unstable 
mode has $\omega_I>0$ with instability timescale $\tau_{\rm{inst}} = 1/\omega_I$. At infinity, we impose that the perturbation is a purely outgoing wave, i.e.,
\begin{equation}\label{asymp_inf}
    \psi(r) \sim e^{i \omega r_*} \,, \quad \text{as} \ r_* \to \infty \,.
\end{equation}
At the object's radius, we impose the boundary conditions in Eqs.~\eqref{eq : axial main bc} and~\eqref{eq : polar main bc} for axial and polar perturbations, respectively.
In the following, we discuss the differences from the BH QNM spectrum for different values of the membrane parameters and the object's compactness.

\subsubsection{Numerical methods for the QNMs}

In this work, we employ the direct integration and continued fractions methods for the numerical computation of the QNMs~\cite{Pani:2013pma, Leaver:1985ax,Pani:2009ss}.
In the direct integration method, the perturbation equations are integrated numerically starting from the membrane location, where the boundary conditions in Eqs.~(\ref{eq : axial main bc}) and~(\ref{eq : polar main bc}) are imposed, to a sufficiently large value of $r$. Then at infinity, the purely outgoing boundary condition in Eq.~\eqref{asymp_inf} is imposed. The latter condition is satisfied  for certain values of the frequency, which are the QNM frequencies. The equations are solved numerically using root-finding algorithms, along with an initial guess for the frequency.
This approach needs to use high-order asymptotic solutions sufficiently far from the compact object (at $r\rightarrow\infty$) and at the horizon of BHs ($r\rightarrow 2M$) to guarantee numerical accuracy. A limitation of the direct integration method is that is does not give accurate results for overtones, and is mainly useful for computing fundamental mode frequencies.

The continued fractions method was originally developed in Ref.~\cite{Leaver:1985ax} to compute the QNMs as roots of implicit equations. The eigenfunction is written as a series expansion whose coefficients satisfy a finite-term recurrence relation.
This method was generalized to work with  boundary conditions at the radius of compact objects in Refs.~\cite{Pani:2009ss,Maggio:2020jml}.

\subsubsection{Validity of the linear-in-spin approximation}
\label{subsec : bh spin lim}

Let us compute the fundamental QNM of a Kerr BH at linear order in spin, and compare it with the fundamental QNM of a Kerr BH as a function of the spin. This comparison allows one to set the maximum spin to which the linear-in-spin approximatiom can be trusted. 
Since axial and polar modes are isospectral in BHs regardless of the spin, we restrict our analysis to the axial QNMs.

To compute the QNM frequencies for a BH to linear order in spin, we impose
either a purely ingoing boundary condition at the horizon as in Eq.~\eqref{horizonasympt} or the membrane boundary condition in Eq.~(\ref{eq : axial main bc}) in the BH limit, i.e., $\epsilon\rightarrow 0$, $\eta\rightarrow \eta_{\rm{BH}}$ and $\zeta\rightarrow \zeta_{\rm{BH}}$. From the numerical integration, $\epsilon=0$ leads to a coordinate singularity, however it is sufficient to choose a sufficiently small value of $\epsilon$, for example $\epsilon = 10^{-10}$, for which the integration is well defined. We checked that the BH QNM does not depend on the choice of the numerical value for $\epsilon$. 

Since our analysis is valid in the slow-spin approximation, we expand the numerically computed QNM frequencies as~\cite{Pani:2013pma,Pierini:2022eim}
\begin{alignat}{3}
\omega(\chi) = \omega^{(0)}+ m \chi \omega^{(1)}+\mathcal{O}(\chi^2),
\label{eq : lin spin qnm}
\end{alignat}
and we discard the second-order-in-spin terms. The numerically solved QNMs even using the linearised boundary conditions have a general dependence on $\chi$. Thus, we Taylor-expand the numerically computed QNMs and truncate them to linear order in spin. The quantity $\omega^{(0)}$ is computed when $\chi=0$, and the quantity $\omega^{(1)}$ is computed as the slope of the linear-in-spin expansion 
by simply using the finite difference formula 
\begin{alignat}{3}
\omega^{(1)} = \frac{\omega(\chi_2)- \omega(\chi_1)}{m(\chi_2-\chi_1)},
\end{alignat}
where $\chi_1 =0$ and we  choose $\chi_2=10^{-4}$. 
Alternatively, we can construct a polynomial fit for $\omega(\chi)$ and truncate the fitted polynomial to the linear order in spin. Both methods yield the same result for $\omega^{(1)}$ within $10^{-3}\%$.

\begin{figure}[t]
\centering
\includegraphics[width=0.45\textwidth]{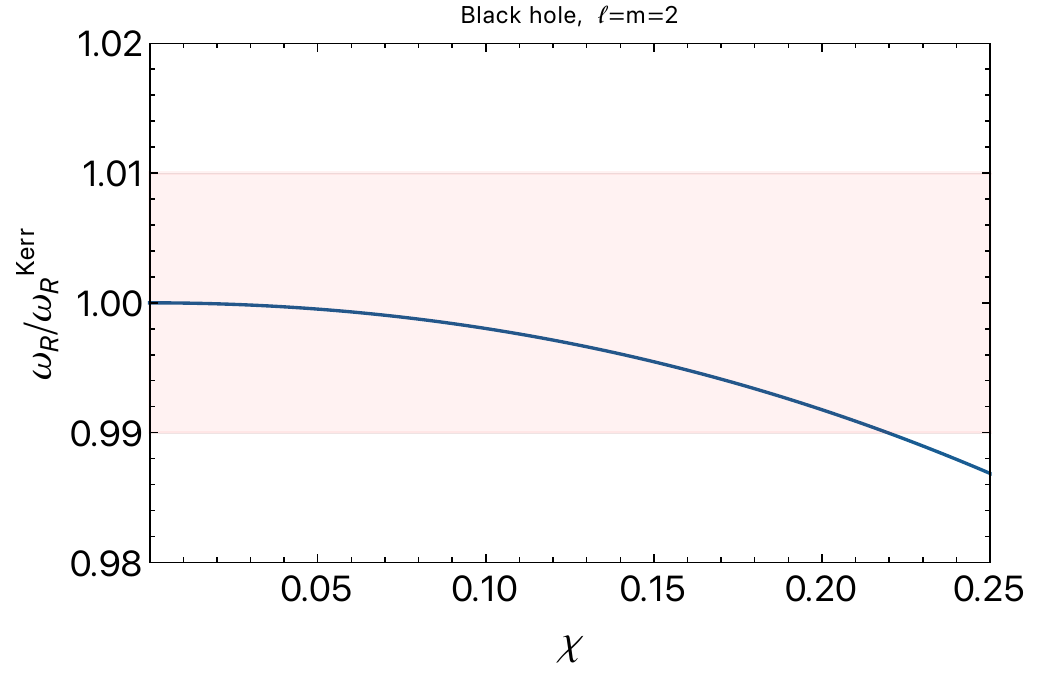}
\includegraphics[width=0.45\textwidth]{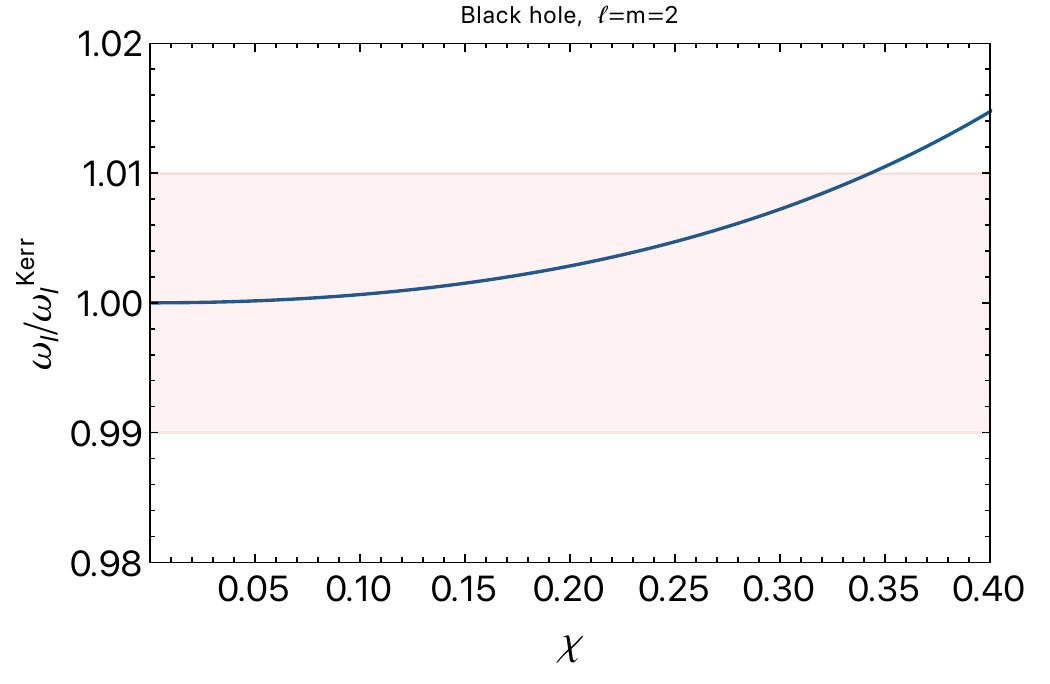}
\caption{Real (top panel) and imaginary (bottom panel) part of the fundamental $\ell=m=2$ QNM of a slowly-spinning BH with respect to a Kerr BH as a function of the spin.
The shaded band shows the 1\% range beyond which the linear-in-spin approximation is not valid. 
}
\label{plot : max spin bh}
\end{figure}
To estimate the maximum value of the spin for which the linear-in-spin approximation is valid, we plot the ratio $\omega/\omega^{\text{Kerr}}$ as a function of the spin in Fig.~\ref{plot : max spin bh}.
The plot shows that the linear-in-spin BH QNMs are consistent with the Kerr QNMs within $1\%$ (shaded region) up to $\chi\sim 0.2$ for the real part of the frequency and $\chi\sim 0.35$ for the imaginary part of the frequency. Based on this argument, we restrict to $\chi\leq 0.2$ as the region in which the linear-in-spin approximation is valid in the rest of this work.

An additional constraint applies if clear unphysical features appear in the results. For instance, as discussed in Sec.~\ref{subsec : obs refl}, the linear-in-spin approximation can lead to spurious superradiant effects for sufficiently small frequencies, i.e., $m\chi/(4M\omega)>1$. In the next section, we show that certain modes become unstable in purely reflecting ultracompact objects with large values of the spin. This is explained as an ergoregion instability, which is a spurious effect at linear order in spin~\cite{Pani:2013pma,Maggio:2018ivz}.

\subsubsection{QNMs with the membrane paradigm}
\label{subsubsec : QNMmem}
Let us now investigate the QNM frequencies of slowly spinning horizonless compact objects, and their variation with the membrane parameters. 

First, we analyse a horizonless compact object with $\epsilon \ll 1$, whose reflectivity is related to the shear viscosity of the membrane as in Fig.~\ref{fig:reflaxial}. In this limit, the axial boundary condition approaches the expression in Eq.~(\ref{eq : axial eps0 lim}). We derive numerically the variation of the QNM frequencies with the shear viscosity $\eta$. This corresponds to the study of how the fundamental mode evolves by varying the reflectivity of the compact object away from the BH case.
\begin{figure}[t]
\centering
\includegraphics[width=0.45\textwidth]{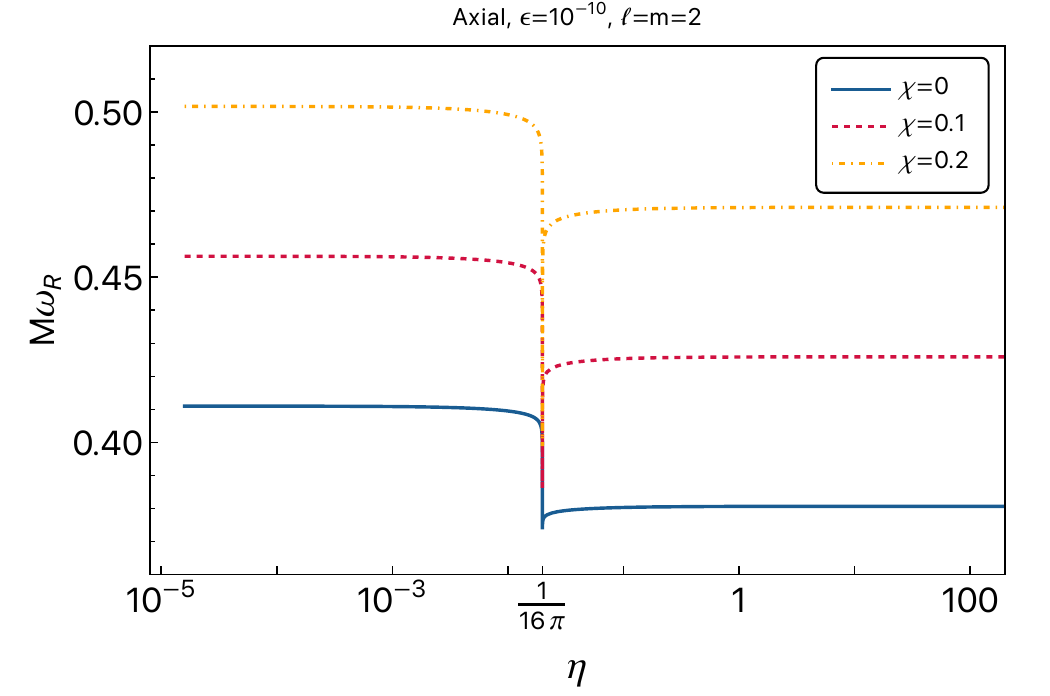}
\includegraphics[width=0.45\textwidth]{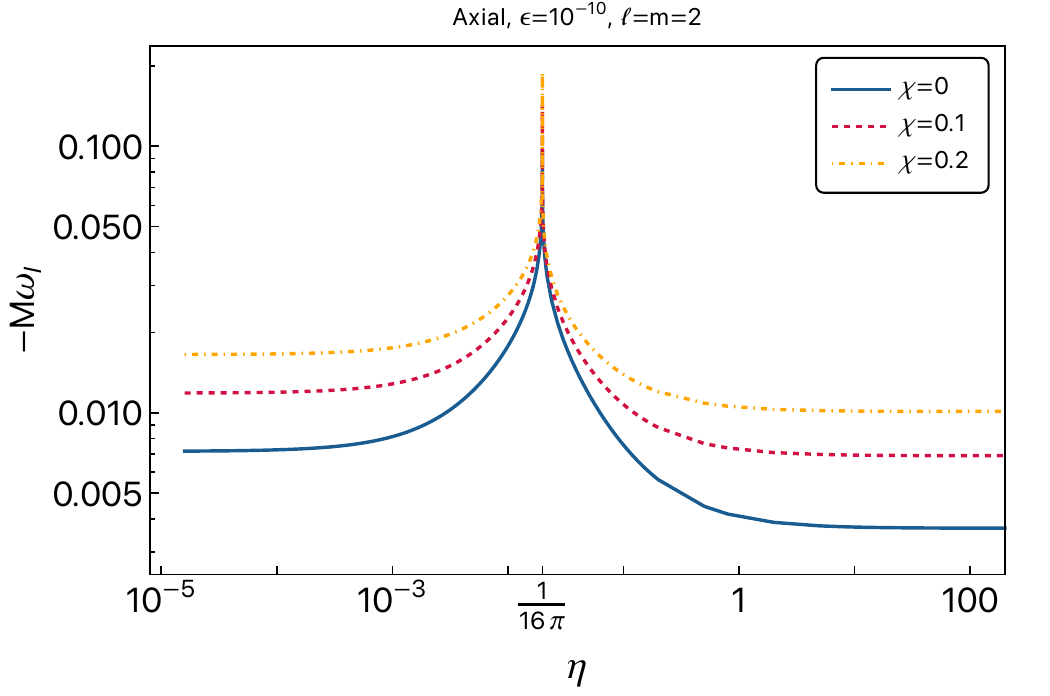}
\caption{Real (top panel) and imaginary (bottom panel) part of the fundamental $\ell=m=2$ QNM of an 
ultracompact object as a function of the viscosity parameter $\eta$ for different values of the spin. For $\eta=1/(16 \pi)$, the compact object is totally absorbing; whereas for $\eta \to 0$ and $\eta \to \infty$, the compact object is perfectly reflecting.}
\label{plot : eta vary axial}
\end{figure}
Fig.~\ref{plot : eta vary axial} shows the real (top panel) and imaginary (bottom panel) part of the axial QNMs of an ultracompact object as a function of the shear viscosity for different values of the spin.
The nonspinning case was derived earlier in Ref.~\cite{Maggio:2020jml}.

When $\eta=1/(16 \pi)$, the compact object is totally absorbing and we recover the BH fundamental QNM.
Then, the variation of the viscosity affects drastically the mode, with a sharp behaviour around the the BH limit. For $\eta \to 0$ and $\eta \to \infty$, the compact object is perfectly reflecting, as shown in Fig.~\ref{fig:reflaxial}.

\begin{figure}[t]
\centering
\includegraphics[width=0.45\textwidth]{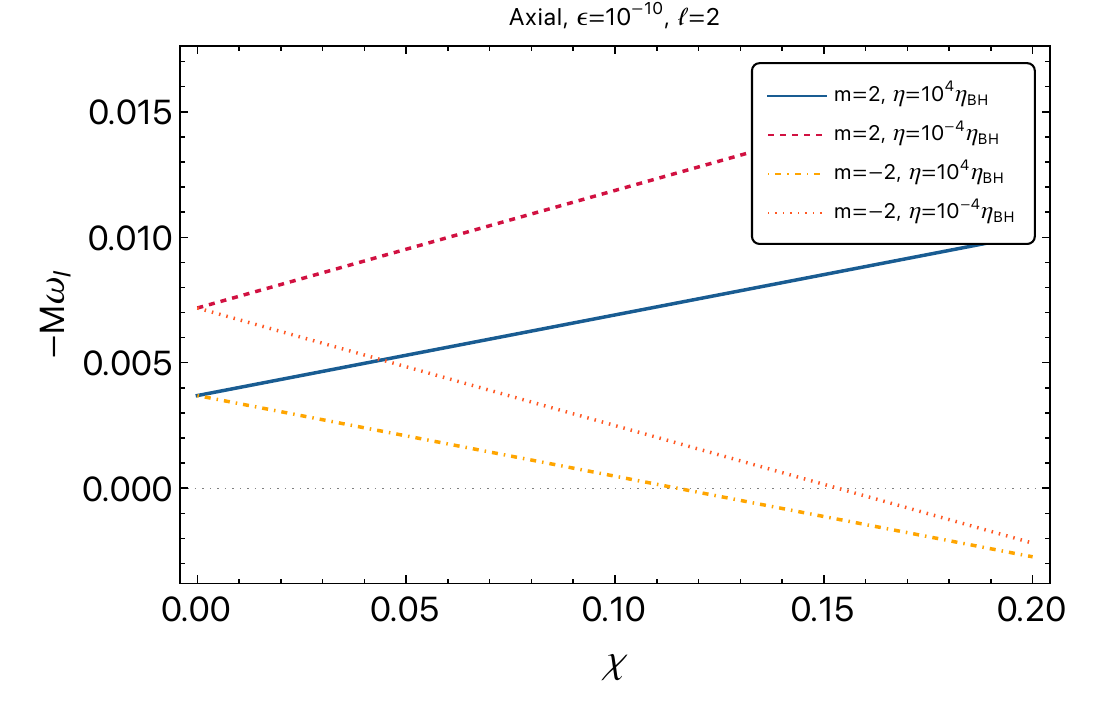}
\caption{Imaginary part of the $\ell=2$, $m=\pm2$ QNMs of an ultracompact object as a function of the spin for a perfectly reflecting compact object with viscosity $\eta\gg\eta_{\rm{BH}}$ and $\eta\ll\eta_{\rm{BH}}$. Note that the imaginary part for $m=-2$ changes sign above a critical value of the spin rendering the mode unstable.
}
\label{plot : extreme eta axial spin}
\end{figure}

It is interesting to note the increase in the imaginary part of the QNMs when the spin increases in the limits $\eta\rightarrow 0,~\infty$. This shows that the $\ell=m=2$ mode decays faster with the spin. Moreover, according to Eq.~(\ref{eq : lin spin qnm}), the $m=-2$ modes should become longer lived. In particular, the imaginary part of these modes can cross 0 indicating an instability (this phenomenon was already observed in Refs.~\cite{Maggio:2018ivz,Zhong:2022jke} for totally reflecting boundary conditions). 
Fig.~\ref{plot : extreme eta axial spin} shows
the imaginary part of the fundamental $(\ell,m)=(2,\pm2)$  QNMs as a function of the spin for the  limits of $\eta \to 0$ and $\eta \to \infty$, corresponding to total reflection. 

We note that the imaginary part of the $m=-2$ QNMs indeed changes sign, signalling an instability in the totally reflecting limits. The critical value of the spin where this happens ($\chi \sim 0.11, 0.15$ for $\eta \to 0, \infty$, respectively) is compatible to Ref.~\cite{Maggio:2018ivz}, where the instability is associated with the ergoregion instability  for spinning and totally reflective ultracompact objects. However, from the earlier discussion in Sec.~\ref{subsec : obs refl}, we know that the presence of ergoregion-related behaviour is an artefact of the expansion at the linear order in spin.
Thus, it is merely a coincidence that the linear-in-spin expansion makes the result more consistent with the behaviour for arbitrary values of the spin. Nevertheless, it is clear from Fig.~\ref{plot : extreme eta axial spin} that the modes 
become longer lived as the spin increases in the purely reflecting limit.

\begin{figure}[t]
\label{plot : axialvpolar}
\centering
\includegraphics[width=0.45\textwidth]{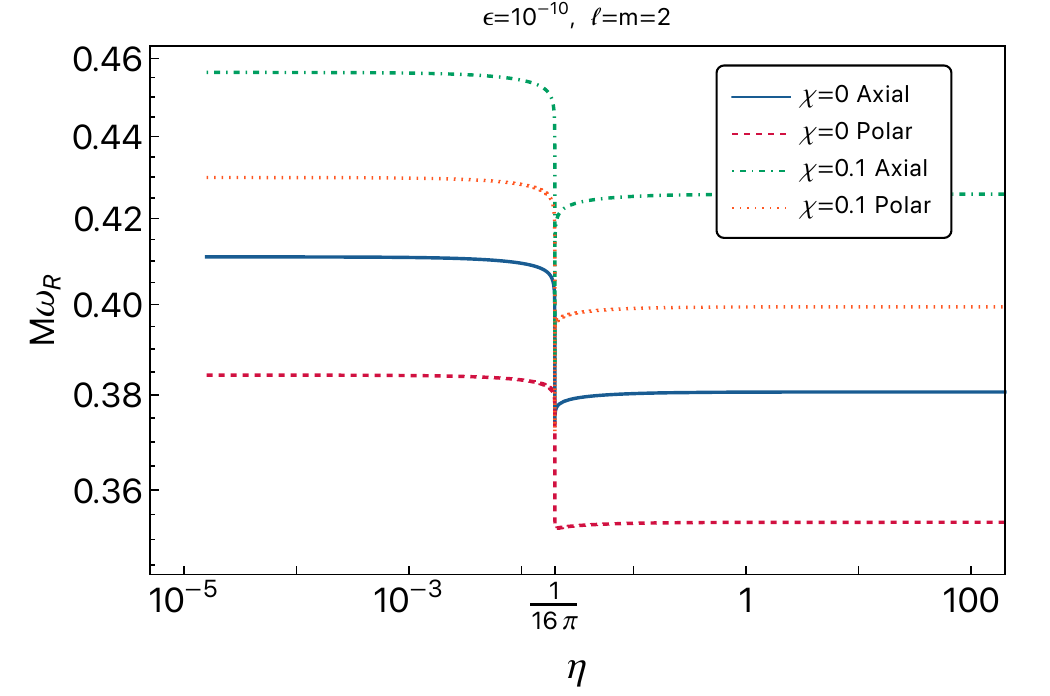}
\includegraphics[width=0.45\textwidth]{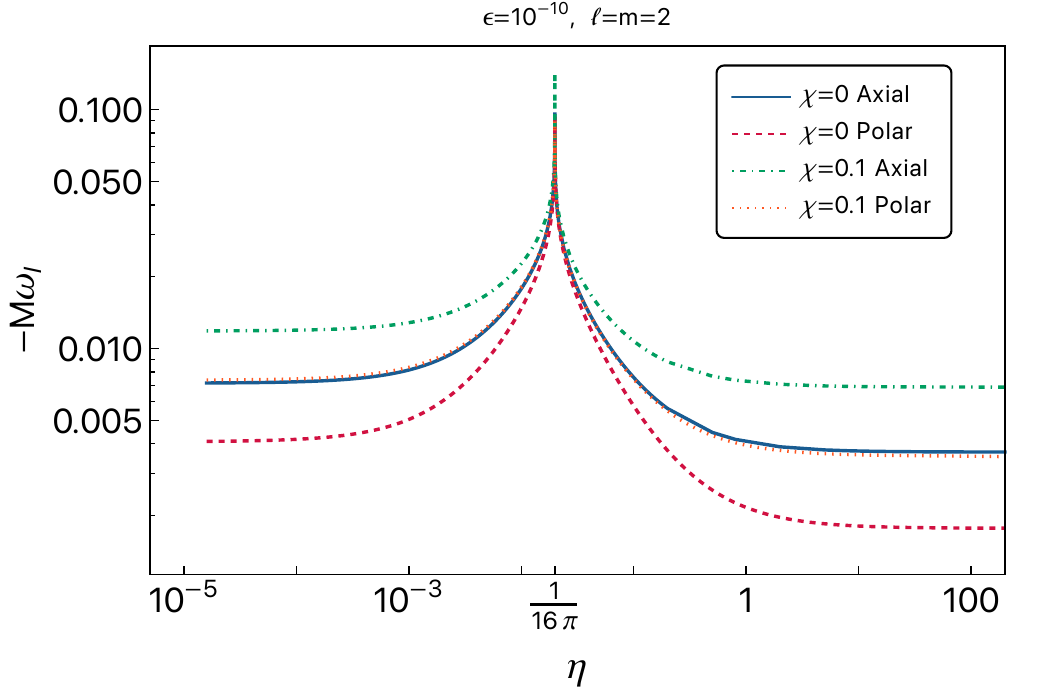}
\caption{Real (top panel) and imaginary (bottom panel) part of the $\ell=m=2$ QNM of an 
ultracompact object as a function of the viscosity parameter $\eta$ for axial and polar perturbations, and different values of the spin. 
Note that the isospectrality between axial and polar modes is broken in both the spinless and spinning cases.
}
\label{plot : polvsaxeta}
\end{figure}

The variation of the polar fundamental QNM mode as a function of the shear viscosity is qualitatively similar to the axial case. Thus, we do not show it here. However, due to the differing boundary conditions in the $\epsilon\rightarrow 0$ limit in Eqs.~(\ref{eq : axial eps0 lim}) and~(\ref{eq : polar eps0 lim}), the isospectrality between axial and polar QNMs is broken when $\eta$ differs from the BH value in both spinning and nonspinning cases. We also note that, for $\epsilon \to 0$, the polar boundary condition does not depend on the bulk viscosity of the membrane. 
Fig.~\ref{plot : polvsaxeta} shows the breaking of isospectrality between axial and polar modes for $\chi=0, 0.1$ as a function of the shear viscosity of the fluid.

\begin{figure}[t]
\includegraphics[width=0.45\textwidth]{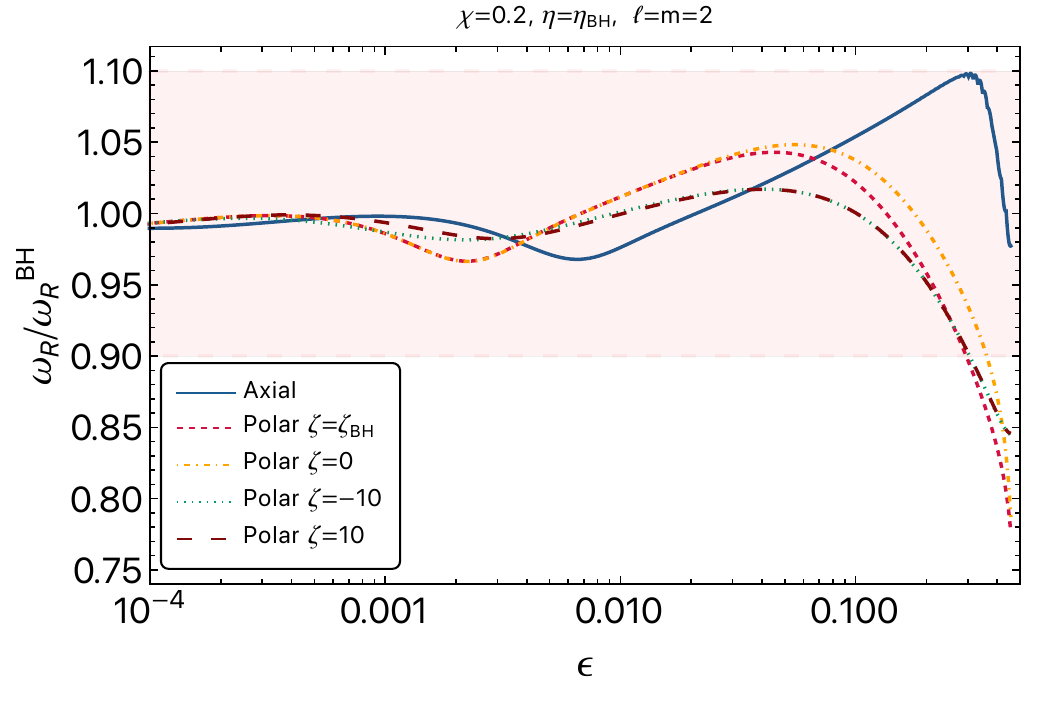}
\includegraphics[width=0.45\textwidth]{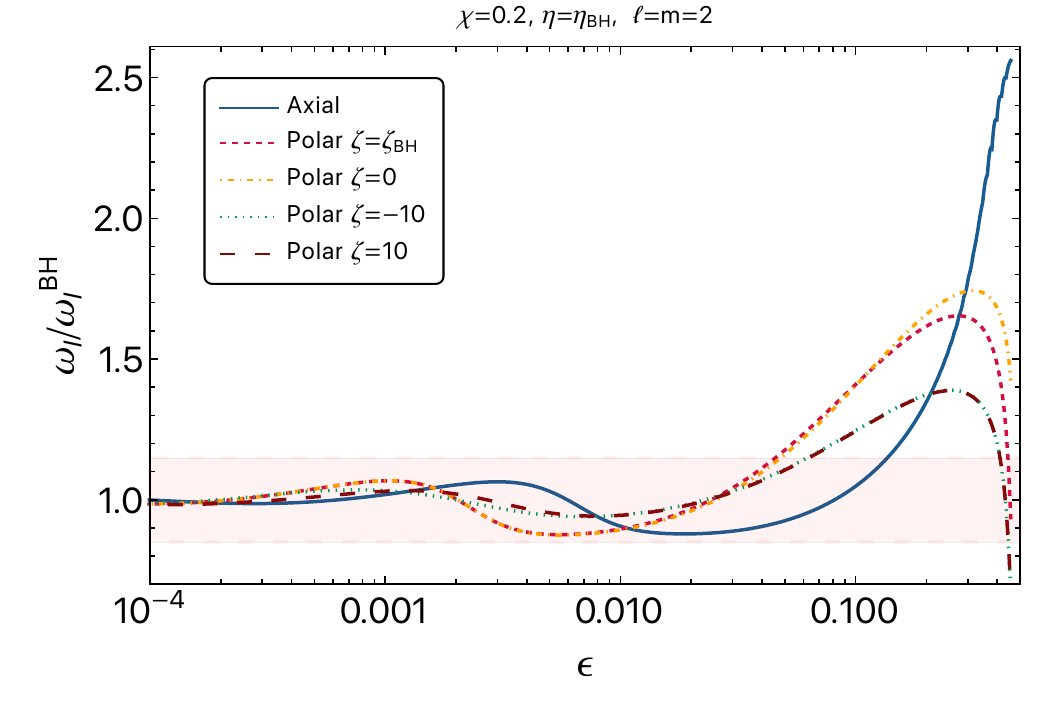}
\caption{
Real (top panel) and imaginary (bottom panel) part of the fundamental $\ell = m=2$ QNM of a compact object with effective shear viscosity $\eta = \eta_{\rm{BH}}$, with respect to their corresponding value for the BH case. The QNMs are functions of the parameter $\epsilon$, which is related to the compactness of the object. For $\epsilon \to 0$, we recover the fundamental BH QNM and axial and polar modes are isospectral. The shaded band correspond to the error bars (10\% and 15\%, respectively) for the real and the imaginary part of the fundamental QNM for the merger event GW150914~\cite{LIGOScientific:2016lio}.}
\label{plot : eps vary spin0d1}
\end{figure}

Let us now analyse the variation of the QNMs with the compactness of the object for a fixed shear viscosity. Fig.~\ref{plot : eps vary spin0d1} shows the real (top panel) and imaginary (bottom panel) part of the axial and polar QNMs of a spinning horizonless compact object as a function of $\epsilon$, where $\chi=0.2$ and $\eta=\eta_{\text{BH}}$. 
For $\epsilon \to 0$ we recover fundamental $\ell = m=2$ QNM of a Kerr BH at the first order in spin. For larger values of $\epsilon$, the compactness of the object decreases and the QNMs depart from the BH picture. The axial and polar QNMs are not isospectral; moreover, the polar QNMs depend on the bulk viscosity of the membrane fluid.

The variation of the fundamental QNM in the spinless case for axial and polar modes was derived in Ref.~\cite{Maggio:2020jml}. It was estimated 
that horizonless compact objects with $\epsilon \lesssim 0.1$ are compatible with the measurement accuracy of the fundamental QNM in GW150914~\cite{LIGOScientific:2016lio}, which is approximately 10\% and 15\% for the real and imaginary part of the frequency, respectively (corresponding to the shaded area in Fig.~\ref{plot : eps vary spin0d1}).
Here, we reconsider this argument by including the spin at the linear order.
By comparing Fig.~\ref{plot : eps vary spin0d1} with Ref.~\cite{Maggio:2020jml}, we find that the variation of the real part of the fundamental mode as a function of $\epsilon$ is suppressed by the spin. However, the variation of the imaginary part of the fundamental mode is enhanced by the spin and it crosses the shaded band for smaller values of $\epsilon$ than in the nonspinning case. 
Fig.~\ref{plot : eps vary spin0d1} shows that spinning ($\chi=0.2$) horizonless compact objects with $\epsilon \lesssim 0.05$ are compatible with measurement accyracy of the fundamental QNM in GW150914.
This suggests that spinning horizonless compact objects may be more easily differentiated than nonspinning ones in the prompt ringdown. Moreover, the breaking of isospectrality between axial and polar modes may be more readily seen for increasing the values of the compactness.

\section{Summary and conclusions}
\label{end}
 In this work, we have extended the membrane paradigm to linear order in spin, to  investigate slowly spinning horizonless compact objects. We have derived the boundary conditions describing axial and polar perturbations of generic compact objects to linear order in spin, and presented their expressions in the Appendix and the supplemental Mathematica document~\cite{githublink}. We have discussed the effect of spin in observables such as the object's reflectivity at infinity and the fundamental QNM frequencies in the ringdown. In particular, we  numerically computed the reflectivity as a function of the shear $\eta$ and bulk $\zeta$ viscosities of the membrane and the object's compactness. We discussed the relation between the viscosity and the reflectivity in the limit of the BH compactness ($\epsilon\rightarrow 0$), where the limits $\eta\rightarrow 0,~\infty$ lead to the cases of purely reflecting compact objects. In the spinning case, the reflectivity changes discontinuously when the object's radius approaches the light ring ($\epsilon\rightarrow 1/2$) even for small values of the spin, both for axial and polar perturbations. 

We analysed the variation of the fundamental QNM frequency with the membrane parameters in the presence of spin. We computed the fundamental $\ell=m=2$ mode for the membrane parameters of a Kerr BH at linear order in spin ($\epsilon\rightarrow 0$, $\eta=1/16\pi$, $\zeta=-1/16\pi$) as a function of the spin, and compared it with the fundamental QNM of a Kerr BH. We found that the linear-in-spin QNM frequency is within 1\% of the Kerr value up to $\chi\sim 0.2$. We thus restricted the upper limit on the spin in this work at $\chi=0.2$. However, considering the presence of spurious superradiance at linear order in spin, we cautioned that one must keep a look out for such unphysical effects, and setting $\chi<0.2$ as the regime of validity for generic compact objects can lead to incorrect conclusions. 

We analysed the shift in the fundamental QNM frequency starting from the BH case ($\eta=1/16\pi$) and going towards the purely reflecting cases ($\eta\rightarrow 0,~\infty$). We noticed that for $m=-2$, the fundamental mode frequencies become longer lived (i.e., the imaginary part tends to 0) as the spin increases in the purely reflecting limits ($\eta\rightarrow 0,~\infty$). In particular, for $\chi\sim 0.1-0.15$, the imaginary part of the fundamental mode switches sign signalling an instability. While this is consistent with previous literature on QNMs of purely reflecting spinning objects, we argued that the linear-in-spin approximation cannot exhibit such effects lacking an ergosphere. Thus, we cannot claim the presence of an instability from the linear-in-spin approach, but the trend of longer lived fundamental mode as the spin is increased for purely reflecting bodies can be taken as a result. We then considered the variation of the fundamental mode frequency as the compactness approaches the BH value, for both axial and polar modes. The polar modes have an additional dependence on the bulk viscosity $\zeta$ when $\epsilon\neq 0$. Similarly to the nonspinning case analysed earlier, we found that the axial and polar modes are not isospectral when $\epsilon\neq 0$. Moreover, the spin enhances the variation of the imaginary part of the fundamental mode with respect to the BH value. Therefore, spinning
horizonless compact objects may be more easily differentiated than nonspinning ones in the prompt ringdown.

Overall, the inclusion of spin causes the QNM spectrum of a horizonless compact object to deviate more strongly from the one of a BH with the same spin parameter, even at linear order when the spin-induced multipole moments are neglected. 
However, the remnants of compact binary coalescences have large values of the spins, $\chi \gtrsim 0.6$. Therefore, it is important to extend this framework to higher orders in spin as a future work.
Finally, the development of a Bayesian framework to map current constraints from parametrised tests of
GR into constraints on the properties of compact objects is left for future work.

\section*{Acknowledgements}
We thank Hector O. Silva and Sebastian Völkel for useful discussions. 
E.M. acknowledges funding from the Deutsche Forschungsgemeinschaft (DFG) - project number: 386119226. 
E.M. is supported by the European Union’s Horizon Europe research and innovation programme under the Marie Skłodowska-Curie grant agreement No. 101107586.
\begin{widetext}
\appendix

\section{Membrane paradigm at the linear order in spin: unperturbed background}
\label{app:background}
The unperturbed metric outside the membrane to linear order in spin is given by :
\begin{equation}
\label{appeq: bg metric}
ds^2\Big|_{(0)} = g^{(0)}_{\mu\nu}dx^{\mu}dx^{\nu} = -f(r) dt^2 + \frac{1}{f(r)}dr^2+ r^2d\Sigma^2 -4 \frac{J}{r}\sin^2\theta dt d\phi,
\end{equation}
where $f(r)=1-2M/r$, 
$d\Sigma^2=d\theta^2+\sin^2\theta d\phi^2$, and 
$J=Ma=M^2\chi$ is the angular momentum of the 
compact object (being modelled by the membrane). The unperturbed membrane is located at the radius $r=r_0$. 
We will use $x^\mu=\{t,r,\theta,\phi\}$ to refer to the global coordinates and $y^a=\{T=t,\Theta=\theta,\Phi=\phi\}$ for the membrane coordinates. The basis vectors for the membrane coordinates are defined as $e_a^{\mu}=\partial x^{\mu}/\partial y^a$, where
\begin{alignat}{3}
\label{appeq: basis0}
e_{T}^{\mu}=\{1,0,0,0\},~e_{\Theta}^{\mu}=\{0,0,1,0\},~e_{\Phi}^{\mu}=\{0,0,0,1\}.
\end{alignat}
The normal vector to the membrane, $n^\mu$, is  fixed by the requirements $n_\mu e_a^{\mu}=0$ and $g^{\mu\nu}n_\mu n_\nu = 1$ to be $n_\mu=\{0,1/\sqrt{f(r)},0,0\}$. The induced metric on the membrane $h_{ab}$ is then given by
\begin{alignat}{3}
    \label{appeq: indmet0}
    ds^2\Big|_{\text{membrane}}=h_{ab}dy^a dy^b=g_{\mu\nu}e_a^{\mu}e_b^{\nu}dy^a dy^b=-f(r_0) dT^2 + r_0^2d\Theta^2+r_0^2\sin^2\Theta d\Phi^2 -4 \frac{J}{r_0}\sin^2\Theta dTd\Phi \,.
\end{alignat}
The extrinsic curvature is defined as $K_{ab}=e_a^{\mu}e_b^{\nu}\nabla_{(\mu}n_{\nu)}$ and is given by
\begin{alignat}{3}
    \label{appeq: K0}
    K_{ab}=\sqrt{f(r_0)}\begin{bmatrix}
-\frac{M}{r_0^2} & 0 & \frac{J\sin^2\Theta}{r_0^2}\\
0 & r_0 & 0\\
\frac{J\sin^2\Theta}{r_0^2} & 0 &  r_0 \sin^2\Theta
\end{bmatrix},
    \end{alignat}
where the trace is $K=K_{ab}h^{ab}=\frac{2r_0-3M}{\sqrt{f(r_0)}r_0^2}$. Thus the left hand side of the junction condition in Eq.~(\ref{junction}) is given by
\begin{alignat}{3}
K_{ab}-Kh_{ab} = \begin{bmatrix}
\frac{2 f(r_0)^{3/2}}{r_0} & 0 & \frac{J(5 r_0-8 M)\sin^2\Theta}{\sqrt{f(r_0)}r_0^3}\\
0 & \frac{M-r_0}{\sqrt{f(r_0)}} & 0\\
\frac{J(5 r_0-8 M)\sin^2\Theta}{\sqrt{f(r_0)}r_0^3} & 0 & \frac{M-r_0}{\sqrt{f(r_0)}}\sin^2 \Theta
\end{bmatrix}.
\label{appeq : Kzero}
\end{alignat}
To relate the extrinsic curvature to the membrane fluid parameters, we also need to compute the energy-momentum tensor in Eq.~(\ref{stress}). This depends on the fluid velocity $u^a$ in membrane coordinates. Intuitively, we expect the fluid to be rotating in the global coordinates, while remaining static on the membrane surface. Thus, we expect the radial and polar components to vanish $u^{\mu} = \{u^t,0,0,J v^\phi\}$, where we factor out $J$ from the azimuthal component of the fluid velocity. Using the metric in Eq.~(\ref{appeq: bg metric}), we can see this is also true for the contravariant components $u_{\mu}=\{u_t,0,0,J v_\phi\}$. We also expect the components to be independent of $t$ and $\phi$ due to stationarity and axial symmetry. In the membrane coordinates, we can write 
\begin{equation}
u_a=u_\mu e_a^\mu=\{u_T=u_t(\Theta),0,u_\Phi=J v_\phi(\Theta)\} \,.
\end{equation}
Now, demanding normalization $u^\mu u_\mu = -1$, we obtain
$u^t=1/\sqrt{f(r)}$ and for the contravariant components, 
\begin{eqnarray}
u_T &=& g_{t\mu}u^{\mu} = g_{tt}u^t + \mathcal{O}(J^2)\approx -\sqrt{f(r_0)} \,, \\
u_{\Phi} &=& g_{\phi\mu}u^{\mu}=-2 J \sin^2\theta/(r_0\sqrt{f(r_0)})+r_0^2  v^\phi)+\mathcal{O}(J^2) \,.
\end{eqnarray}
Plugging this into Eq.~(\ref{stress}), we obtain the stress energy tensor to be 
\begin{alignat}{3}
-8 \pi T_{ab} = -8 \pi\begin{bmatrix}
\rho f(r_0) & 0 & \frac{J\sin^2\Theta}{r_0}\left[2 \rho - f(r_0)r_0^3(\rho+p)v^\phi\right]\\
0 & p r_0^2 & -r_0^2 J \sin^2\Theta \eta \partial_\Theta v^\phi\\
\frac{J\sin^2\Theta}{r_0}\left[2 \rho - f(r_0)r_0^3(\rho+p)v^\phi\right] & -r_0^2 J \sin^2\Theta \eta \partial_\Theta v^\phi & p r_0^2\sin^2\Theta
\end{bmatrix}.
\label{appeq : zero stress}
\end{alignat}
Now, comparing Eq.~(\ref{appeq : zero stress}) with Eq~(\ref{appeq : Kzero}), and imposing the junction condition in Eq.~(\ref{junction}), we get 
\begin{alignat}{3}
\rho=-\frac{\sqrt{f(r_0)}}{4\pi r_0},~ p =\frac{1+f(r_0)}{16 \pi r_0 \sqrt{f(r_0)}},~J v^\Phi=\frac{2J}{r_0^3\left(1-3f(r_0)\right)\sqrt{f(r_0)}} \,.
\end{alignat}
The membrane fluid parameters with $J \to 0$ are consistent with the ones obtained in the spinless case in Ref.~\cite{Maggio:2020jml} .

\section{Perturbation equations}
\label{app : pert eqn}

 The perturbations can be expanded in the spherical harmonic basis as 
\begin{alignat}{3}
g_{\mu\nu}&=g_{\mu\nu}^{0}+\epsilon_p\delta g_{\mu\nu}, \label{appeq : metpert} \\
 \nonumber \delta g_{\mu\nu}& =\sum_{\ell,m} \Big\{\left[ H_0^{\ell m}(r,t) dt^2 + H_2^{\ell m}(r,t) dr^2 + r^2 K^{\ell m}(r,t)d\Sigma^2  \nonumber +2 H_1^{\ell m}(r,t) dr dt\right]Y_{\ell m}(\theta,\phi) + \\& 2 h_0^{\ell m}(r,t) 
 \left[S_\theta^{\ell m}(\theta,\phi) d\theta + S_\phi^{\ell m}(\theta,\phi) d\phi \right] dt + 2 h_1^{\ell m}(r,t) \left[ S_\theta^{\ell m}(\theta,\phi) d\theta + S_\phi^{\ell m}(\theta,\phi) d\phi \right] dr \Big\},
\label{appeq : met pert}
\end{alignat}
where $S_{\theta}^{\ell m}(\theta,\phi) =-\partial_{\phi}Y_{\ell m}(\theta,\phi)/\sin\theta$,  $S_{\phi}^{\ell m}(\theta,\phi) =\sin\theta\partial_{\theta}Y_{\ell m}(\theta,\phi)$, $\epsilon_p$ is a small parameter controlling the strength of perturbation, and we restrict to linear order in $\epsilon_p$. The metric perturbation can be split into the axial and polar sectors considering their transformation under parity. In the absence of spin, perturbations with different parity do not mix, and satisfy the Regge-Wheeler and Zerilli equations for axial and  polar perturbations, respectively. To the linear order in spin, the coupling between axial and polar modes can be neglected as far as the computation of the QNM frequencies is concerned.
\subsection{Axial sector}
\label{app : pert eqn axial}
The axial sector is the part of the metric perturbation that transforms as $(-1)^{\ell+1}$ under parity transformation, i.e., ($\theta\rightarrow \pi-\theta$, $\phi\rightarrow \phi+\pi$). This is given by
\begin{alignat}{3}
\delta g_{\mu\nu}^{\text{axial}}  & = 2 \sum_{\ell,m} \left\{ h^{\ell m}_{0}(r,t)\left[S_{\theta}^{\ell m}(\theta,\phi) d\theta +  S_{\phi}^{\ell m}(\theta,\phi)d\phi \right]dt + h_{1}(r,t) \left[S_{\theta}^{\ell m}(\theta,\phi) d\theta + S_{\phi}^{\ell m}(\theta,\phi)d\phi \right]dr \right\}.
\end{alignat}
The Einstein equations in vacuum, $G_{\mu\nu}=0$, for axial perturbations to the linear order in spin and in the perturbation, can be reduced to a modified Regge-Wheeler equation to first order in spin given by \cite{Pani:2013pma}
\begin{alignat}{3}
& \frac{d^2\psi_{\text{RW}}(r)}{d r_*^2} + \left[\omega^2 - \frac{4 m J \omega}{r^3} -f(r)\left(\frac{\ell(\ell+1)}{r^2}- \frac{6 M}{r^3}  + \frac{24 J m (-7 M + 3 r) }{\ell(\ell+1) r^6 \omega}\right)\right] \psi_{\text{RW}}(r) = 0,
\end{alignat}
where
\begin{alignat}{3}
\psi_{\text{RW}}(r)=h_1(r)\frac{f(r)}{r}\left(1+\frac{2m J}{r^3 \omega}\right),
\end{alignat}
and we have suppressed the $\ell,m$ indices for simplicity of notation. The other function in the axial sector, $h_0(r)$, is related to $h_1(r)$ through the relation
\begin{alignat}{3}
h_0(r)=\frac{i f(r)\left[\ell(\ell+1)r^4\omega h_1(r)f'(r)+f(r)(-12J q m h_1(r)+\tilde{q}r^4\omega h_1'(r)\right]}{\tilde{q}\omega r(-2J m + r^3\omega)},
\label{app eq h1toh0}
\end{alignat}
where $q=(\ell-1)(\ell+2)/2$, $\tilde{q}=\ell(\ell+1)$, and the prime stands for radial derivative ($d/dr$).
Note that in the above equations, we have factorized the time-dependence as $h_1(r,t)=h_1(r)e^{-i\omega t}$ and similarly for $h_0(r,t)$.
\subsection{Polar sector}
The polar sector transforms as $(-1)^\ell$ under parity transformations, and is given by
\begin{alignat}{3}
\delta g_{\mu\nu}^{\text{polar}} = & \sum_{\ell,m}\left[H_0^{\ell m}(r,t) dt^2 + H_2^{\ell m}(r,t)dr^2 + r^2 K(r,t) d\Sigma^2  +2  H_1^{\ell m}(r,t)dr dt \right] Y_{\ell m}(\theta,\phi).
\end{alignat}
The Einstein equations for the polar perturbations can be reduced to a modified Zerilli equation to the first order in spin given by~\cite{Pani:2013pma}
\begin{alignat}{3}
\frac{d^2 \psi_{\text{Z}}(r)}{d r_*^2} + \left[\omega^2 - \frac{4 m J \omega}{r^3} -f(r)V_{\text{Z}}(r)\right] \psi_{\text{Z}}(r) = 0,
\end{alignat}
where
\begin{alignat}{3}
\label{appeq : Zer pot}
& V_{\text{Z}}(r) = \frac{2 M}{r^3}+\frac{ (\ell -1) (\ell +2)}{3} \left[\frac{2 (\ell -1) (\ell +2) (\ell ^2+\ell +1)}{(6 M+r (\ell ^2+\ell -2))^2}+\frac{1}{r^2} \right] + \\& \frac{4 m J}{r^7 \omega  \ell  (\ell +1) (6 M+r (\ell ^2+\ell -2))^4} \Big\{27648 M^6+2592 M^5 r (6 \ell  (\ell +1)-19)+144 M^4 r^2 \Big[6 r^2 \omega ^2+\ell  (\ell +1) \times \nonumber\\&(21 \ell  (\ell +1)-148)+230 \Big]  +12 M^3 r^3 (\ell ^2+\ell -2) \Big[72 r^2 \omega ^2+\ell  (\ell +1) (29 \ell  (\ell +1)-200)+374\Big]+ 12 M^2 r^4 \times \nonumber \\& (\ell ^2+\ell -2)^2  \Big[28 r^2 \omega ^2+\ell  (\ell +1) (5 \ell  (\ell +1)-12)-4\Big]  +M r^5 (\ell ^2+\ell -2)^2 \Big[24 r^2 \omega ^2 (2 \ell  (\ell +1)-5)+(\ell -1) \times \nonumber \\& (\ell +2) (\ell ^2+\ell +2) (7 \ell  (\ell +1)-38)\Big]+r^6 (\ell ^2+\ell -2)^3 \Big[2 r^2 \omega ^2 (\ell ^2+\ell -4)  -3 (\ell ^2+\ell -2) (\ell ^2+\ell +2)\Big]\Big\}. \nonumber
\end{alignat}
The Zerilli function $\psi_{\text{Z}}(r)$ is related to the radial functions in the polar sector via the relations
\begin{alignat}{3}
\label{appeq : Zer func1}
\begin{bmatrix}
    K(r) \\
    H_1(r)
\end{bmatrix} 
= 
\begin{bmatrix}
    A_{11}  &  A_{12} \\
    A_{21}  & A_{22}
\end{bmatrix} 
. 
\begin{bmatrix}
    \tilde{\psi}_{\text{Z}}(r) \\
    \tilde{\psi}_{\text{Z}}'(r)
\end{bmatrix},
\end{alignat}
where 
\begin{alignat}{3}
\label{appeq : Zer func2}
\mathbf{A} = \begin{bmatrix}
    \frac{12 M^2 + 6 M q r + \ell(\ell+1) q r^2}{2 r^2 (3 M + q r)} & 1 \\
    -i\omega \left[1+M \left(\frac{1}{2M-r}- \frac{6}{2M+2 q r}\right)\right] & \frac{i\omega r^2}{2M -r}
\end{bmatrix} \,,
\end{alignat}
and 
\begin{alignat}{3}
\label{appeq : Zer func3}
\tilde{\psi}_{\text{Z}}(r) = \psi_{\text{Z}}(r)\left[1+ \frac{2 m J (12 M^3 - 6 M^2 r + q r^3(q+r^2\omega^2)+M(-2 q^2 r^2 + 3 r^4 \omega^2)}{\ell(\ell+1)r^4(3M+ q r)^2\omega}\right].
\end{alignat}
The other functions describing polar metric perturbrations, $H_0$ and $H_2$, can be written in terms of  $H_1$ and $K$ via the relations
\begin{alignat}{3}
H_2(r) &= \frac{r^2}{(r-2M)^2} H_{0}(r) + 4 m J \omega r \frac{K(r)}{\ell(\ell+1)(r-2M)^2}, \label{H2H0K} \\ 
 H_0(r) &= \frac{i(2M-r)(\tilde{q} M - 2 r^3 \omega^2)}{2 r^2 (3M + q r)\omega} H_1(r) - \frac{3 M^2 + M(-1 + 2 q)r - q r^2 + r^4 \omega^2)}{r(3M + q r)}K(r) \label{H0H1K} \\& \nonumber + \frac{m J}{2\tilde{q}r^5(3M + qr)^2\omega^2} \{i(2M-r) \left[\tilde{q}^2 M (3M+qr)-2 (\tilde{q}M- 2 r^3\omega^2)(\tilde{q}M+ r^3\omega^2)\right]H_1(r) \\& + r\omega [4(\tilde{q} M + r^3 \omega^2)(3 M^2+M(-1+ 2 q)r - q r^2 + r^4 \omega^2) + (6 M + 2 q r)(2(3 M - 2r ) r^3\omega^2 \nonumber \\& + \tilde{q} (-3 M^2 + M r + 2 r^4 \omega^2))]K(r) \}. \nonumber
\end{alignat}
Although the Zerilli equation appears different than the Regge-Wheeler equation, they both yield the same spectrum of QNMs when the boundary condition is that of BHs. This can be shown in the non-spinning case via the Chandrashekar transformation, and at the linear order in spin with a more general transformation.
For a generic compact object described by the membrane paradigm, the boundary condition at the membrane surface can break the isospectrality.

\section{Membrane paradigm at the linear order in spin: gravitational perturbations}
\label{app : mem bc}
In accordance with the junction conditions, the extrinsic curvature and the induced metric from either side of the membrane surface must satisfy
\begin{alignat}{3}
(K^+_{ab} - h^+_{ab} K^+)-(K^-_{ab} - h^-_{ab} K^-) = - 8 \pi T_{ab}, \qquad h^+_{ab}  = h^{-}_{ab}, 
\end{alignat}
where $T_{ab}$ is the surface stress-energy tensor of the membrane. Normally, to proceed further, one needs to fix the spacetime in the interior region to the membrane. However, in this work, the membrane is intended to be an effective way to parametrise an arbitrary compact object. This  choice does not break any symmetries and lets the stress-energy tensor do all the heavy lifting.  A simple way to implement this choice is to demand $K^{-}_{ab}=0$, which is also the standard choice when the membrane is used to describe BHs. Then, the junction condition reduces to 
\begin{alignat}{3}
K_{ab} - h_{ab} K = - 8 \pi T_{ab}, 
\label{appeq: special jc}
\end{alignat}
where we have removed the $+$ signs denoting that  quantities are evaluated outside the membrane for convenience. The surface stress energy tensor is given by \cite{Maggio:2020jml}
\begin{alignat}{3}
T_{ab} = \rho u_au_b + (p-\zeta\Theta)\gamma_{ab}- 2 \eta \sigma_{ab}, 
\end{alignat}
where $\Theta = h^{ab}\nabla_{b}u_{a}$, is the expansion rate, $\sigma_{ab} = \nabla_{(a} u_{b)}-\frac{1}{2}\Theta \gamma_{ab} $ is the shear tensor, and $\gamma_{ab} = h_{ab}+u_au_b$ is the projection tensor orthogonal to the fluid velocity. The derivation of the density, pressure and fluid velocity when the membrane is unperturbed is given in Appendix~\ref{app:background}. Here, we derive the boundary conditions for the metric perturbations at the surface of the membrane to the linear order in spin.

When the perturbation is added to the background, its contribution appears both on the membrane parameters and the metric perturbations in Eq.~(\ref{appeq: special jc}). Thus, the outside metric is given by Eq.~(\ref{appeq : metpert}), and the membrane surface is parametrised as
\begin{alignat}{3}
x^{\mu}(y^a)&= \left(t=T, r=r_0+\epsilon_p \delta r(\Theta,\Phi) e^{-i\omega T}, \theta=\Theta,\phi=\Phi\right), \\
\delta r(\Theta,\Phi) &= \sum_{\ell,m}\xi^{\ell m}_rY_{\ell m}(\Theta,\Phi), \label{appeq : rad_sft}
\end{alignat}
where $(T, \Theta, \Phi)$ are intrinsic coordinates to parametrise the (perturbed) membrane, and $\delta r(\Theta,\Phi)$ is the radial deformation of the membrane surface. We can derive the perturbed tangent vectors as $e_a^{\mu}=dx^\mu/dy^a=(e_a^{\mu})^{0}+\epsilon_p \delta e_a^{\mu}$ where
\begin{alignat}{3}
\delta e_{a}^{\mu} \equiv \begin{bmatrix} \delta e_t^{\mu}  \\ \delta e_\Theta^{\mu} \\ \delta e_\phi^{\mu}\end{bmatrix} = \sum_{\ell m}\begin{bmatrix} 0 & -i\omega \xi_r^{\ell m} Y_{\ell m}(\Theta,\Phi)& 0 & 0\\ 0 & \xi^{\ell m}_r \partial_\Theta Y_{\ell m}(\Theta,\Phi) & 0 & 0 \\ 0  & \xi_r^{\ell m} \partial_\Phi Y_{\ell m}(\Theta,\Phi) & 0 & 0\end{bmatrix} e^{-i\omega T} \,,
\end{alignat}
and the basis vectors in the unperturbed case, $(e_a^\mu)^{0}$, is given in Eq.~(\ref{appeq:  basis0}). We can now derive the perturbations in the induced metric $h_{ab}=g_{\mu\nu}e_a^{\mu}e_{b}^{\nu}=h_{ab}^{0}+\epsilon_p\delta h_{ab}e^{-i\omega T}$, with $h_{ab}^{0}$ given in Eq.~(\ref{appeq: indmet0}) and 
\begin{alignat}{3}\delta h_{ab}=
\sum_{\ell m}\begin{bmatrix} 
\frac{r_0^2 H_0^{\ell m}(r_0)-2 M \xi_{r}^{\ell m}}{r_0^2} & -\csc\Theta h^{\ell m}_0(r_0)\partial_\Phi & h^{\ell m}_0(r_0)\sin\Theta \partial_\Theta +\frac{2J\sin^2 \Theta}{r_0^2}\xi_r^{\ell m}\\ -\csc\Theta h^{\ell m}_0(r_0)\partial_\Phi  & r_0 \left(r_0 K^{\ell m}(r_0)+2 \xi_r^{\ell m}\right) & 0 \\ h_0^{\ell m}(r_0)\sin\Theta \partial_\Theta +\frac{2J\sin^2\Theta}{r_0^2}\xi_r^{\ell m} & 0 & r_0 \left(r_0 K^{\ell m}(r_0)+2 \xi_r^{\ell m}\right)\sin^2\Theta\end{bmatrix}
Y_{\ell m}(\Theta,\Phi).
\end{alignat}
The normal vector $n_{\mu}=n_{\mu}^{0} + \epsilon_p \delta n_{\mu} e^{-i\omega T}$, defined through the relations $n_\mu e_a^{\mu}=0$, and $n_\mu n_\nu g^{\mu\nu}=1$, is perturbed as
\begin{alignat}{3}
\delta n_{\mu} =\sum_{\ell m} \begin{bmatrix} i\omega\frac{\xi_r^{\ell m}}{\sqrt{f(r_0)}}  \\ \frac{H_2^{\ell m}(r_0)}{2}\sqrt{f(r_0)} \\ -\frac{\xi_r^{\ell m} }{\sqrt{f(r_0)}}\partial_\Theta\\ -\frac{\xi_r^{\ell m}}{\sqrt{f(r_0)}}\partial_\Phi
\end{bmatrix}Y_{\ell m}(\Theta,\Phi).\,,
\end{alignat}
where $n_\mu^{0}=\{0,1/\sqrt{f(r_0)},0,0\}$. Subsequently, we can compute the perturbation to the extrinsic curvature $K_{ab}=K_{ab}^{0}+\epsilon_p\delta K_{ab} e^{-i\omega T}$, using $K_{ab}=e_a^{\mu}e_b^{\nu}\nabla_{(\mu}n_{\nu)}$. $K_{ab}^{0}$ is defined in Eq.~(\ref{appeq: K0}), and
\begin{alignat}{3}
\delta K_{TT} &= \sum_{\ell m}\frac{\left(-5 f^2+6 f+4 r_0^2 \omega ^2-1\right) \xi _r^{\ell m}+r_0 f \left(-f^2 H_2^{\ell m}+f H_2^{\ell m}+4 i \omega r_0 H_1^{\ell m}+2 r_0 (H_0^{\ell m})'\right)}{4 r_0^2 \sqrt{f}} Y_{\ell m}(\Theta,\Phi) \,,\nonumber   
\\ \delta K_{\Theta\Theta} \nonumber &=\sum_{\ell m}\frac{f \left(2 \partial_\Phi h_1^{\ell m} \csc \Theta \left(\cot \Theta-\partial_\Theta\right)+r_0^2 (K^{\ell m})'+2 r_0 K^{\ell m}+\xi _r^{\ell m}\right)-r_0 f^2 H_2^{\ell m}+\left(1-2 \partial_\Theta^2\right) \xi _r^{\ell m}}{2 \sqrt{f}}Y_{\ell m}(\Theta ,\Phi) \,,
\\ \nonumber \delta K_{T\Theta}&=\sum_{\ell m} \left[\frac{f \left[\partial_\Theta H_1^{\ell m}+i \partial_\Phi \csc \Theta \left(\omega  h_1^{\ell m}+i (h_0^{\ell m})'\right)\right]+2 i \omega  \partial_\Theta \xi _r^{\ell m}}{2 \sqrt{f}}+\frac{2 J \cot \Theta \left( f h_1^{\ell m} \sin \Theta \partial_\Theta-\partial_\Phi \xi _r^{\ell m}\right)}{r_0^3 \sqrt{f}} \right] Y_{\ell m}(\Theta,\Phi) \,, \label{appeq: K1}
\\ \nonumber \delta K_{T\Phi}&=\sum_{\ell m} \left\{ \frac{f \left(i \omega   h_1^{\ell m} \sin \Theta \partial_\Theta+ \sin \Theta  \partial_\Theta (h_0^{\ell m})'-\partial_\Phi H_1^{\ell m}\right)+2 i \omega  \partial_\Phi \xi _r^{\ell m}}{2 \sqrt{f}} + \right. \\ &\nonumber \left. \frac{J \left[-f \left(4 \partial_\Phi h_1^{\ell m} \cos \Theta +5 \sin ^2\Theta \xi _r^{\ell m}\right)+r_0 f^2 H_2^{\ell m} \sin ^2\Theta +\sin \Theta \xi _r^{\ell m} \left(4  \cos \Theta \partial_\Theta+\sin \Theta \right)\right]}{2 r_0^3 \sqrt{f}} \right\} Y_{\ell m}(\Theta,\Phi ) \,,
\\\nonumber  \delta K_{\Theta\Phi}&=\sum_{\ell m}\frac{2 \partial_\Phi \left(\cot \Theta -\partial_\Theta\right) \xi _r^{\ell m}-f h_1^{\ell m} \sin \Theta  \left(\partial_\Phi^2 \csc ^2\Theta -\partial_\Theta^2+ \cot \Theta \partial_\Theta\right)}{2 \sqrt{f}} Y_{\ell m}(\Theta ,\Phi ) \,,
\\ \delta K_{\Phi\Phi} &= \delta K_{\Theta\Theta}\sin^2 \Theta+ \sum_{\ell m}\frac{2 \partial_\Phi f h_1^{\ell m} \left( \sin \Theta \partial_\Theta-\cos \Theta \right)-\xi _r^{\ell m} \left(-\sin ^2\Theta \partial_\Theta^2 + \sin \Theta  \cos \Theta \partial_\Theta+\partial_\Phi^2\right)}{\sqrt{f}}Y_{\ell m}(\Theta ,\Phi ) \,, 
\end{alignat}
where we have suppressed the radial argument in the perturbation variables for simplicity of notation. All the radial functions and their derivatives are to be evaluated at $r=r_0$. Also note that $\partial_\Theta$  ($\partial_\Phi$) denote a differentiation with respect to $\Theta$ ($\Phi$), acting on the spherical harmonic $Y_{\ell m}(\Theta,\Phi)$, and the prime denotes a radial derivative. 

The fluid membrane parameters are also perturbed so that Eq.~(\ref{appeq: special jc}) continues to hold. We parametrise the  fluid perturbations as 
\begin{alignat}{3}\label{fluidpert}
\rho&=\rho^{0}+\epsilon_p\delta \rho(\Theta,\Phi)e^{-i\omega T},\\
p&=p^{0}+\epsilon_p\delta p(\Theta,\Phi)e^{-i\omega T}, \\
u^{a}&=(u^{a})^{0}+\epsilon_p\delta u^a(\Theta,\Phi)e^{-i\omega T}.
\end{alignat}
Furthermore, we expand the membrane perturbations in spherical harmonics (analogous to stellar perturbation theory in Ref.~\cite{Kojima:1992ie}) as
\begin{alignat}{3}
\delta u^{\Theta}&= \sum_{\ell m}\Big[V^{\ell m} \frac{S^{\ell m}_{\phi}(\Theta,\Phi)}{\sin \Theta}+U^{\ell m}  S^{\ell m}_\theta(\Theta,\Phi)\Big],
\\ \delta u^{\Phi}&= \sum_{\ell m}\Big[-V^{\ell m} \frac{S^{\ell m}_\theta(\Theta,\Phi)}{\sin \Theta}+U^{\ell m} S^{\ell m}_\phi(\Theta,\Phi)\Big], \\
\delta \rho(\Theta,\Phi) &= \sum_{\ell m}\delta\rho^{\ell m}Y_{\ell m}(\Theta,\Phi),\\
\delta p(\Theta,\Phi)&=\sum_{\ell m}\delta p^{\ell m}Y_{\ell m}(\Theta,\Phi).
\end{alignat}
This makes the process of deriving the boundary conditions  more convenient, and splits the perturbations based on their transformation under parity. 
The perturbed stress energy tensor is then given by
\begin{eqnarray}
\delta T_{TT} &=& \sum_{\ell m} \left[ \frac{\sqrt{f} \left((f-1) \xi_r^{\ell m}+r_0 H_0^{\ell m}\right)}{4 \pi r_0^2}+f \delta \rho ^{\ell m}-\frac{J  ( \sin \Theta U^{\ell m}\partial _{\Theta }+\partial_\Phi U^{\ell m})}{4 \pi  r_0^2} \right] Y_{\ell m}(\Theta ,\Phi ),\nonumber 
\\ 
\delta T_{\Theta\Theta} &=& \sum_{\ell m}\Big\{r_0^2 \Big[\text{$\delta p$}^{\ell m}-2 \eta  \partial_{\Phi } \cot \Theta \csc \Theta U^{\ell m}+ \left(2 \eta  \partial_{\Phi } \csc \Theta U^{\ell m}-(\zeta -\eta ) \cot \Theta  V^{\ell m }\right)\partial_{\Theta }- \nonumber  \\
&~&  V^{\ell m}\Big(\partial_{\Phi }^2 (\zeta -\eta ) \csc ^2 \Theta +\partial_{\Theta }^2 (\zeta +\eta ) \Big) \nonumber\Big]+ \frac{\left(r_0 K^{\ell m}+2 \xi _r^{\ell m}\right) \left(-M+8 i \pi  \zeta  r_0^2 \omega +r_0\right)}{8 \pi  r_0 \sqrt{f}} + \\
&~& \frac{J \left[-\partial_{\Phi } \left(2 \zeta  r_0 K^{\ell m}+(3 \zeta +\eta ) \xi _r^{\ell m}\right)-6 i r_0 \omega   (\zeta -\eta ) h_0^{\ell m} \sin \Theta \partial_{\Theta }\right]}{2 r_0 \sqrt{f} \left(r_0-3 M\right)} \nonumber +\frac{J(\zeta-\eta)}{f^{\frac{3}{2}}(3f-1)r_0^2} \times \\
&~& \Big[4 i r_0 \omega   h_0^{\ell m} \sin \Theta \partial_{\Theta }-\partial_{\Phi } \left(\xi _r^{\ell m}-r_0 H_0^{\ell m}\right)-6 i r_0^3 \omega  f^{3/2} \left(\sin \Theta  U^{\ell m}\partial_{\Theta } +\partial_{\Phi } V^{\ell m}\right)+4 i r_0^3 \omega  \sqrt{f} \times \nonumber \\
&~& \left( \sin \Theta  U^{\ell m}\partial_{\Theta }+\partial_{\Phi } V^{\ell m}\right)\Big]\Big\}Y_{\ell m} (\Theta ,\Phi ) \,, \nonumber \\
\delta T_{T\Theta}&=&\nonumber\sum_{\ell m} \Big\{\frac{r_0^2 \partial_{\Theta } \left(3 f-1\right) V^{\ell m}-\partial_{\Phi } \csc \Theta \left(4 \sqrt{f} h_0^{\ell m}+3 r_0^2 f U^{\ell m}-r_0^2 U^{\ell m}\right)}{16 \pi  r_0} -\\
&~& \frac{2 \eta  J \sin \Theta  \left[ -U^{\ell m}\partial_{\Theta }^2+\partial_{\Phi } \csc ^2 \Theta  \left(\partial_{\Phi } U^{\ell m}+2 \cos \Theta  V^{\ell m}\right)+ \csc \Theta \left(\cos \Theta  U^{\ell m}\partial_{\Theta }-2 \partial_{\Phi } V^{\ell m}\right)\right]}{r_0 \left(3 f-1\right)} \Big\} Y_{\ell m}(\Theta,\Phi ) \,, \nonumber 
\\ 
\delta T_{T\Phi}&=&\nonumber\sum_{\ell m} \Big\{ \frac{r_0^2 \partial_{\Phi } \left(3 f-1\right) V^{\ell m}- \sin \Theta \partial_{\Theta } \left(4 \sqrt{f} h_0^{\ell m}+3 r_0^2 f U^{\ell m}-r_0^2 U^{\ell m}\right)}{16 \pi  r_0}  - \frac{J }{r_0 \left(r_0-3 M\right)} \times \\
&~&\nonumber \Big[6 M \sin ^2 \Theta  \delta \rho ^{\ell m}+r_0 \Big(\sin ^2 \Theta  (\text{$\delta p$}^{\ell m}-\delta \rho ^{\ell m})+2 \eta  \partial_{\Phi } \cos \Theta  U^{\ell m}+ \sin \Theta  \left(2 \eta  \partial_{\Phi } U^{\ell m} +(\zeta +\eta ) \cos \Theta  V^{\ell m}\right)\partial_{\Theta } \\
&~& -(\zeta -\eta ) \sin ^2 \Theta  V^{\ell m}\partial_{\Theta }^2 -\partial_{\Phi }^2 (\zeta +\eta ) V^{\ell m}\Big)\Big]
 +\frac{J\sin ^2 \Theta  }{16 \pi  r_0^3 f^{3/2} \left(3 f-1\right)}\Big[-i r_0 f \left(-3 i H_0^{\ell m}+32 \pi  \zeta  r_0 \omega  K^{\ell m}\nonumber \right.\\
&~& \left.-2 i K^{\ell m}+64 \pi  \zeta  \omega  \xi _r^{\ell m}\right)+f^2 \left(6 r_0 K^{\ell m}\nonumber+\xi _r^{\ell m}\right)+24 f^3 \xi _r^{\ell m}+r_0 H_0^{\ell m}-\xi _r^{\ell m}\Big] \Big\} Y_{\ell m}(\Theta,\Phi)  \,, \\
\delta T_{\Theta\Phi}&=&\sum_{\ell m} \Big\{ \eta  r_0^2 \csc \Theta  \left[ -U^{\ell m}\sin ^2 \Theta  \partial_{\Theta }^2+\partial_{\Phi } \left(\partial_{\Phi } U^{\ell m}+2 \cos \Theta  V^{\ell m}\right)+\sin \Theta  \left(\cos \Theta  U^{\ell m}-2 \partial_{\Phi } V^{\ell m}\right)\partial_{\Theta } \right] + \nonumber\\
&~& \nonumber \frac{J \sin \Theta  }{8 \pi  r_0^2 f^{3/2} \left(3 f-1\right)} \Big[f \left(8 \pi  \eta   \sin \Theta  \partial_{\Theta }\xi _r^{\ell m}+3 \partial_{\Phi } h_0^{\ell m} \left(3+16 i \pi  \eta  r_0 \omega \right)\right)-9 \partial_{\Phi } f^2 h_0^{\ell m}-3 i r_0^2 f^{3/2} \left(16 \pi  \eta  r_0 \omega -3 i\right) \\
&~&  \left( \sin \Theta  V^{\ell m}\partial_{\Theta }-\partial_{\Phi } U^{\ell m}\right)+2 r_0^2 \sqrt{f} \left(1+16 i \pi  \eta  r_0 \omega \right) \left( \sin \Theta  V^{\ell m}\partial_{\Theta }-\partial_{\Phi } U^{\ell m}\right) +9 r_0^2 f^{5/2} \times \nonumber \\
&~& \nonumber \left( \sin \Theta  V^{\ell m}\partial_{\Theta }-\partial_{\Phi } U^{\ell m}\right) -2 \partial_{\Phi } h_0^{\ell m} \left(1+16 i \pi  \eta  r_0 \omega \right)+8 \pi  \eta  r_0  H_0^{\ell m} \sin \Theta \partial_{\Theta }-8 \pi  \eta   \sin \Theta  \xi _r^{\ell m}\partial_{\Theta }\Big] \Big\} Y_{\ell m}(\Theta ,\Phi ),
\end{eqnarray}
\begin{eqnarray}
\delta T_{\Phi\Phi} &=& \delta T_{\Theta\Theta}\sin^2\Theta -\sum_{\ell m}\Big\{ 2 \eta  r_0^2 \left[\partial_{\Phi } \left(\partial_{\Phi } V^{\ell m}-2 \cos \Theta  U^{\ell m}\right)+ \sin \Theta  \left(2 \partial_{\Phi } U^{\ell m}+\cos \Theta  V^{\ell m}\right)\partial_{\Theta }- \sin ^2 \Theta  V^{\ell m}\partial_{\Theta }^2\right] + \nonumber \\
&~& \nonumber \frac{J \sin^2\Theta f}{4 \pi  r_0^2 f^{3/2} \left(3 f-1\right)} \Big[\left(8 \pi  \eta  \partial_{\Phi } \xi _r^{\ell m}-3 i h_0^{\ell m} \sin \Theta  \partial_{\Theta } \left(16 \pi  \eta  r_0 \omega -3 i\right)\right)+9  f^2 h_0^{\ell m} \sin \Theta \partial_{\Theta } -3 i r_0^2 f^{3/2} \\
&~& \nonumber\left(16 \pi  \eta  r_0 \omega -3 i\right) \left( \sin \Theta  U^{\ell m}\partial_{\Theta }+\partial_{\Phi } V^{\ell m}\right)+2 r_0^2 \sqrt{f} \left(1+16 i \pi  \eta  r_0 \omega \right) \left( \sin \Theta  U^{\ell m}\partial_{\Theta }+\partial_{\Phi } V^{\ell m}\right)+9 r_0^2 f^{5/2} \times \\
&~&  \left( \sin \Theta  U^{\ell m}\partial_{\Theta }+\partial_{\Phi } V^{\ell m}\right)+2  h_0^{\ell m} \sin \Theta  \partial_{\Theta }\left(1+16 i \pi  \eta  r_0 \omega \right)-8 \pi  \eta  \partial_{\Phi } \left(\xi _r^{\ell m}-r_0 H_0^{\ell m}\right)\Big] \Big\} Y_{\ell m}(\Theta,\Phi). 
\end{eqnarray}
Now, using Eq.~\eqref{appeq: special jc}, we can write the constraints on the perturbations to these quantities as
\begin{alignat}{3}
\delta K_{ab} - h_{ab}\delta K  - K\delta h_{ab} = -8 \pi \delta T_{ab},
\end{alignat}
which is a total of six equations that fix the perturbations to the membrane as well as the boundary condition. We project the equations onto the spherical harmonics $Y_{\ell m}(\Theta,\Phi)$ and extract the relations for a given ($\ell,m$) mode (and its interaction with the neighbouring modes)~\cite{Kojima:1992ie, Pani:2013pma} yielding to
\begin{alignat}{3}
\label{app eq: 03}
(ab)=(T\Phi)\rightarrow ~ &  \ell(\ell+1)\beta^{\ell m}_0+i m J [(\ell-1)(\ell+2)\chi_0^{\ell m}+\tilde{\alpha}^{\ell m}_{0}+\eta_0^{\ell m}]-J Q^{\ell m}(\ell+1)[(\ell-2)(\ell-1)\xi_0^{\ell-1,m}\nonumber \\&-(\ell-1)\tilde{\beta}_0^{\ell-1,m} +\zeta_0^{\ell-1,m}]  + J Q^{\ell+1,m}\ell[(\ell+2)(\ell+3)\xi_0^{\ell+1,m}+(\ell+2)\tilde{\beta}^{\ell+1,m}_0+\zeta_0^{\ell+1,m}] =0, \\ 
(\Theta\Theta)-(\Phi\Phi)\rightarrow ~ & \ell(\ell+1) K_{00}^{\ell m}+i m J  K_{gg}^{\ell m}-J Q^{\ell m}(\ell+1)K_{ff}^{\ell-1,m}+ J Q^{\ell+1,m}lK_{ff}^{\ell+1,m}=0,
\end{alignat}
for the predominantly axial sector, and similarly
\begin{alignat}{3}
(ab)= (TT)\rightarrow & ~ A^{\ell m}_{00}+i m J C^{\ell m}_0 + J Q^{\ell m}B_0^{\ell-1,m}- J Q^{\ell+1,m}B_0^{\ell+1,m} =0,\\
(\Theta\Theta)+(\Phi\Phi) \rightarrow & ~ A^{\ell m}_{33}+i m J C^{\ell m}_3 + J Q^{\ell m}B_3^{\ell-1,m}- J Q^{\ell+1,m}B_3^{\ell+1,m} =0, \\
(T\Theta) \rightarrow & ~ \ell(\ell+1)\alpha^{\ell m}_0+i m J [(\ell-1)(\ell+2)\xi_0^{\ell m}-\tilde{\beta}^{\ell m}_{0}-\zeta_0^{\ell m}]-J Q^{\ell m}(\ell+1)[(\ell-2)(\ell-1)\chi_0^{\ell-1,m}\nonumber\\& +(\ell-1)\tilde{\alpha}_0^{\ell-1,m} -\eta_0^{\ell-1,m}]  - J  Q^{\ell+1,m}\ell[(\ell+2)(\ell+3)\chi_0^{\ell+1,m}-(\ell+2)\tilde{\alpha}^{\ell+1,m}_0-\eta_0^{\ell+1,m}] =0, 
\\ (\Theta\Phi) \rightarrow & ~\ell(\ell+1) K_{s}^{\ell m}+i m J  K_{f}^{\ell m}- J Q^{\ell m}(\ell+1)K_{g}^{\ell-1,m}+J Q^{\ell+1,m}lK_{g}^{\ell+1}=0,
\end{alignat}
for predominantly polar sector. Here, $Q^{\ell m} = \sqrt{(\ell^2-m^2)/(4l^2-1)}$ and the other variables are functions of the individual modes of axial and polar perturbations which help us to organize the  equations in a similar manner as in Ref.~\cite{Kojima:1992ie,Pani:2013pma}. Note that among the terms proportional to $J$, each equation contains functions that are proportional to $\ell$ and the neighbouring modes $\ell-1$, $\ell+1$. Among these, the latter ones lead to a mixing of axial and polar modes, as parity conservation prevents any coupling between axial and polar modes with the same angular number. For example in Eq.~(\ref{app eq: 03}), $\beta_0^{\ell m}$, $\chi_0^{\ell m}$, $\tilde{\alpha}_0^{\ell m}$ and $\eta_0^{\ell m}$ depend only on the axial perturbation functions, i.e., $h_{0/1/2}^{\ell m}(r_0)$ and $U^{\ell m}$; whereas $\xi_0^{\ell m}$, $\tilde{\beta}_0^{\ell m}$ and $\zeta_0^{\ell m}$, which contribute as neighbouring modes, depend only on the polar perturbation functions, i.e., $\xi_r^{\ell m}$, $V_0^{\ell m}$, $\delta p^{\ell m}$, $\delta \rho^{\ell m}$, $H_{0/1/2}^{\ell m}(r_0)$ and $K^{\ell m}(r_0)$. 

To leading order in spin, we can drop the mixing between axial and polar modes without affecting the derived QNM frequencies. This is true both in the equations of motions and in the junction conditions. To see this, we first note that the equations are invariant under 
\begin{alignat}{3}
 a^{\ell m}\rightarrow (-1)^{\ell+1}a^{\ell,-m},\quad p^{\ell m} \rightarrow (-1)^\ell p^{\ell,-m}, \quad  J \rightarrow -J, \quad m \rightarrow -m,
\end{alignat}
where $a^{\ell m}$ collectively refers to all the linear functions of axial perturbations in the above equations (e.g., $\beta_0^{\ell m}$) and $p^{\ell m}$ collectively refers to all the linear functions of polar perturbations in the above equations (e.g., $A_{00}^{\ell m}$). At the first order in spin, the QNM frequencies can be expanded as $\omega=\omega_0+m\chi \omega_1 + \mathcal{O}(\chi^2)$, where $J=M^2\chi$. This implies that the linear-in-spin terms in the equations of motion that do not have an extra factor of $m$ cannot affect $\omega_1$, and may be dropped for their computation. This is also true for the perturbation equations, as shown in Ref.~\cite{Pani:2013pma}. Thus, we can decouple the axial and polar sectors and simplify the junction conditions to
\begin{alignat}{3}
\label{app eq axialsimp1}
(ab)=(T\Phi)\rightarrow ~ &  \ell(\ell+1)\beta^{\ell m}_0+i m J [(\ell-1)(\ell+2)\chi_0^{\ell m}+\tilde{\alpha}^{\ell m}_{0}+\eta_0^{\ell m}]=0, \\ 
(\Theta\Theta)-(\Phi\Phi)\rightarrow ~ & \ell(\ell+1) K_{00}^{\ell m}+i m J  K_{gg}^{\ell m}=0,
\label{app eq axialsimp2}
\end{alignat}
for axial modes and 
\begin{alignat}{3}
(ab)= (TT)\rightarrow & ~ A^{\ell m}_{00}+i m J C^{\ell m}_0=0,\\
(\Theta\Theta)+(\Phi\Phi) \rightarrow & ~ A^{\ell m}_{33}+i m J C^{\ell m}_3=0, \\
(T\Theta) \rightarrow & ~ \ell(\ell+1)\alpha^{\ell m}_0+i m J [(\ell-1)(\ell+2)\xi_0^{\ell m}-\tilde{\beta}^{\ell m}_{0}-\zeta_0^{\ell m}]=0, 
\\ (\Theta\Phi) \rightarrow & ~\ell(\ell+1) K_{s}^{\ell m}+i m J  K_{f}^{\ell m}=0,
\end{alignat}
for polar modes to the first order in spin. 
\subsection{Axial boundary condition}
Using Eqs.~(\ref{app eq axialsimp1}) and~(\ref{app eq axialsimp2}), we can eliminate the axial matter perturbation ($V^{\ell m}$) between these two equations to get the boundary condition for the Regge-Wheeler function $\psi_{\text{RW}}(r)$ to linear order in spin to be
\begin{alignat}{3}
\label{app eq axialbcfin}
\frac{1}{\psi_{\rm{RW}}(r)}\frac{d\psi_{\rm{RW}}(r)}{d r_*}\Big|_{r=r_0}&= \frac{3 M^2 y^2 \nu V_{\text{RW}}(r_0)-6 i w^3+2 i w^2 y}{6 M w \nu (3 w-y)} \nonumber \nonumber \\&\nonumber \nonumber \frac{m \chi}{6 \ell (\ell+1) M y^5 (y-3 w)^2\nu    }[w^3 (36 w^2 y (i (\ell^2+\ell+9) y+16 \nu +\nu  y^2)+6 w y^2 ((\ell^2+\ell-7) \nu  y^2\\ & -2 i (2 \ell^2+2 \ell+13) y-42 \nu )+y^3 (-3 (\ell^2+\ell-4) \nu  y^2+4 i (\ell^2+\ell+6) y+36 \nu )-216 i w^3 \nonumber \\& (y-2 i \nu ))-3 M^2 w y^4 V_{\text{RW}}(r_0) (4 \nu  w-6 i w y+\nu  y^3+2 i y^2-2 \nu  y)],
\end{alignat}
where $w=M\omega$, $y=\omega r_0$, $\nu=16 \pi \eta$ and
\begin{alignat}{3}
V_{\rm{RW}}(r_0) = f(r)\Big( \frac{\ell(\ell+1)}{r^2} - \frac{6 M}{r^3}  + \frac{24 m J (-7 M + 3 r_0)}{\ell(\ell+1)r_0^6 \omega}\Big ).
\end{alignat}
The above boundary condition is consistent with the earlier results in Ref.~\cite{Maggio:2020jml} in the non spinning case for $\chi=0$. In the BH limit, i.e., $\eta\rightarrow (16 \pi)^{-1}$ and $r_0\rightarrow 2M$, we have
\begin{alignat}{3}
\frac{1}{\psi_{\rm{RW}}(r)}\frac{d\psi_{\rm{RW}}(r)}{d r_*}\Big|_{r=r_0} = - i\Big(\omega - \frac{m \chi}{4 M }\Big) = -i(\omega-m\Omega_{H}),
\end{alignat}
as expected. 
\subsection{Polar boundary condition}
The polar boundary condition requires an extra ingredient, which is the equation of state of the membrane fluid. Since there are four equations in the polar sector and four matter perturbations, i.e., $\delta \rho^{\ell m}$, $\delta p^{\ell m}$, $V^{\ell m}$ and $\xi_r^{\ell m}$, we do not have any remaining constraint upon eliminating the matter variables to serve as the boundary condition. Thus, we additionally impose that the perturbations to the pressure and the density are related via the equilibrium sound-speed \cite{Kojima:1992ie,Maggio:2020jml}
\begin{alignat}{3}
c_s^2 = \frac{\delta p^{\ell m}}
{\delta \rho^{\ell m}}=  \frac{dp_0/dr}{d\rho^0/dr}  = -\frac{3M^2-3M r_0 + r_0^2}{2(6 M^2-5 M r_0+ r_0^2)}.
\end{alignat}
With this extra constraint, we 
obtain the boundary condition to the  linear order in spin for the Zerilli function $\psi_{\text{Z}}(r)$, using the relations in Appendix.~\ref{app : pert eqn}. 
We write it implicitly here as
\begin{alignat}{3}
\label{app eq polarbcfin}
\frac{1}{\psi_{\text{Z}}(r)}\frac{d\psi_{\text{Z}}(r)}{dr_*}\Big|_{r=r_0} = -16 \pi \eta i \omega + G(r_0,\omega,\eta,\zeta) + m \chi H(r_0,\omega,\eta,\zeta). \,,
\end{alignat}
where $G(r_0,\omega,\eta,\zeta)$ is given explicitly in Ref.~\cite{Maggio:2020jml}. We provide the polar boundary condition in the supplemental Mathematica document in~\cite{githublink}. Note that unlike the axial sector, there is dependence on the bulk viscosity both in spinless case and at the linear order in spin for the polar sector. In the BH limit, i.e., $\eta\rightarrow (16 \pi)^{-1}$, $\zeta\rightarrow -(16 \pi)^{-1}$ and $r_0\rightarrow 2M$, we have
\begin{alignat}{3}
\frac{1}{\psi_{\text{Z}}(r)}\frac{d\psi_{\text{Z}}(r)}{d r_*}\Big|_{r=r_0} = - i\Big(\omega - \frac{m \chi}{4 M }\Big) = -i(\omega-m\Omega_{H}),
\end{alignat}
as expected.
\end{widetext}

\bibliography{example.bib}

\end{document}